\documentclass[12pt]{article}%
\usepackage{graphicx}
\usepackage{amsmath}
\usepackage{amsfonts}
\usepackage{amssymb}%
\newtheorem{theorem}{Theorem}

\newcommand{\al}{\alpha}
\newcommand{\be}{\beta}
\newcommand{\ga}{\gamma}

\newcommand{\Om}{\Omega}

\newcommand{\R}{\mathbb R}
\newcommand{\C}{\mathbb C}
\newcommand{\Z}{\mathbb Z}

\begin{document}

\title{\textbf{Gibbs Ensembles of Nonintersecting Paths}}
\author{Alexei Borodin and Senya Shlosman\\Caltech, Pasadena, USA\\borodin@caltech.edu;\\Centre de Physique Theorique, CNRS,\\Luminy, Marseille, France\\shlosman@cpt.univ-mrs.fr\\and IITP, RAS, Moscow, Russia\\shlosman@iitp.ru}
\maketitle

\begin{abstract}
We consider a family of determinantal random point processes on the
two-dimensional lattice and prove that members of our family can be
interpreted as a kind of Gibbs ensembles of nonintersecting paths.
Examples include probability measures on lozenge and domino tilings
of the plane, some of which are non-translation-invariant.

The correlation kernels of our processes can be viewed as extensions
of the discrete sine kernel, and we show that the Gibbs property is
a consequence of simple linear relations satisfied by these kernels.
The processes depend on infinitely many parameters, which are
closely related to parametrization of totally positive Toeplitz
matrices.

\end{abstract}

\section{Introduction}

It is well known that the Gibbs random fields are defined via
prescribing their conditional distributions. In the case of the
nearest-neighbor interactions these distributions are given by some
relatively simple relations. The computation of their correlation
functions is on the other hand usually a very difficult problem,
because it requires the passing to the thermodynamic limit. In
comparison, the determinantal random fields (or random point
processes, both terms are used) are defined in such a way that the
correlation functions are given by relatively simple direct
formulas, while the computation of the conditional distributions may
again require taking the thermodynamic limit, since the dependence
is usually long-range.

The purpose of the present paper is the study of some 2D random
fields $n=\left\{  n_{t}=0,1,t\in\mathbb{Z}^{2}\right\}  ,$ which
are both determinantal and have in addition some kind of the Gibbs
property. We wanted to understand which properties of the kernel
produce the Gibbsianity of the random field. It turns out that the
property sought is some linear relation on the matrix elements of
the kernel. Below we explain this statement for a certain class of
2D determinantal random fields.

Every random field is specified by the (consistent) assignment of the
probabilities $\mathbf{P{}r}$ to the events $\left\{  n_{t_{i}}=1,n_{t_{j}%
}=0\right\}  ,$ for any two non-intersecting finite sets $\left\{
t_{i}\right\}  ,$ $\left\{  t_{j}\right\}  $. We say that a random field $n$
is a determinantal random field with the kernel $K\left(  t^{\prime}%
,t^{\prime\prime}\right)  ,$ if for every finite collection $t_{i}%
\in\mathbb{Z}^{2}$%
\begin{equation}
\mathbf{P{}r}_{K}\left\{  n_{t_{i}}=1\right\}  =\det\left\Vert K\left(
t_{i},t_{j}\right)  \right\Vert .\label{11}%
\end{equation}
The inclusion-exclusion principle then implies that the probability of a more
general event
\[
\mathbf{P{}r}_{K}\left\{  n_{t_{i}}=1,n_{t_{j}}=0\right\}  =\left(  -1\right)
^{h}\det\left\Vert \tilde{K}\left(  t_{i},t_{j}\right)  \right\Vert ,
\]
where
\[
\tilde{K}\left(  t_{i},t_{j}\right)  =\left\{
\begin{array}
[c]{cc}%
K\left(  t_{i},t_{j}\right)\quad   & \text{ if }t_{i}\neq t_{j},\\
K\left(  t_{i},t_{i}\right)\quad   &\quad \text{if }t_{i}=t_{j}\text{ and }n_{t_{i}%
}=1,\\
K\left(  t_{j},t_{j}\right)  -1 &\quad \text{if }t_{i}=t_{j}\text{ and }n_{t_{j}%
}=0,
\end{array}
\right.
\]
and $h$ is the number of \textquotedblleft holes\textquotedblright,
i.e. indices $\left\{  t_{j}:n_{t_{j}}=0\right\}  .$ We will refer
to the sites with values $1$ as \textquotedblleft
particles\textquotedblright. The above formula is sometimes referred
to as \emph{complementation principle}, cf. A.3 of \cite{BOO}.

In this paper we study random fields $n,$ corresponding to the
kernels $K$ constructed as follows. Suppose that for every
$k\in\mathbb{Z}^{1}$ the function $\psi_{k}\left(  u\right)  $ is
given, which can be one of the following four functions:
$$\left(
1-\alpha_{k}^{+}u\right) ^{-1},\quad \left(
1-\alpha_{k}^{-}u^{-1}\right) ^{-1},\quad \left(
1+\beta_{k}^{+}u\right),\quad \left( 1+\beta_{k}^{-}u^{-1}\right) ,
$$
with positive constants $\alpha^{\pm},$ $\beta^{\pm}$. Let us also
fix a complex number $z$ with $\Im z>0$ and denote $C_\pm$ any
contour that joins $\bar{z}$ and $z$ and crosses the real axis at a
point of $\R_{\pm}$.

 For $t^{\prime}=\left( \sigma,x\right)  ,$
$t^{\prime\prime}=\left( \tau,y\right)  $ we define
\begin{equation}\label{kernel}
K\left(  \sigma,x;\tau,y\right)  \equiv K_{\sigma,\tau}\left(  x-y\right)
=
\begin{cases}
\dfrac{1}{2\pi i}\displaystyle\int_{C_+}\left(
\prod_{k=\sigma+1}^{\tau}\psi
_{k}\left(  u\right)  \right)  ^{-1}\frac{du}{u^{x-y+1}} & \text{ if  }%
\sigma<\tau,\\
& \\
\dfrac{1}{2\pi i}\displaystyle\int_{C_+}\frac{du}{u^{x-y+1}} &
\text{if  }\sigma
=\tau,\\
& \\
\dfrac{1}{2\pi
i}\displaystyle\int_{C_-}\prod_{k=\tau+1}^{\sigma}\psi_{k}\left(
u\right) \frac{du}{u^{x-y+1}} & \text{ if  }\sigma>\tau.
\end{cases}
\end{equation}

Our main results concerning such determinantal random fields are threefold:

\begin{enumerate}
\item Due to the (particles \textit{or }holes) interlacing property of our
fields $n,$ they can be interpreted as ensembles of non-intersecting
infinite random lattice paths. (These ensembles are different for
different kernels, and will be described in detail below.)

\item The collections $\mathbf{\omega=}\left\{  \omega_{i}\right\}  $ of
random lattice paths thus obtained are \textquotedblleft Gibbs
random paths ensembles\textquotedblright. They are defined by the
action functional $S_{K}\left(  \omega\right)  ,$ which is local and
is determined by the parameters of the kernel $K$.

\item The validity of the above two statements follows from
 simple linear
relations that the matrix elements of the kernel $K$ satisfy, and do not
depend on $K$ otherwise.
\end{enumerate}

\subsection{\label{Gibbs} Gibbs Path Ensembles}

We will define the Gibbs random path ensemble, corresponding to the
additive functional $S.$ Here $S$ is a function defined on the set
of all finite selfavoiding lattice paths $\omega$, which has the
additivity property: if $\omega=\omega_{1}\cup\omega_{2},$
$\omega_{1}\cap\omega_{2}=\{t\}\in\omega,$ then $S\left(
\omega\right) =S\left(  \omega_{1}\right)  +S\left(  \omega
_{2}\right)  .$ Let $\mu$ be a probability distribution on the set
of families of non-intersecting double-infinite polygonal lattice
paths $\mathbf{\omega =}\left\{  \omega_{i},-\infty<i<\infty\right\}
$. Let $\Lambda\subset \mathbb{Z}^{2}$ be a finite box, i.e. a
finite connected subset of $\Z^2$ with connected complement. Let the
paths $\omega_{i}$ be fixed outside $\Lambda.$ We denote the
restriction of $\mathbf{\omega}$ to the complement of $\Lambda$ by
$\mathbf{\omega}_{\bar{\Lambda}}.$ Some of the paths from
$\mathbf{\omega}$ are entering and exiting $\Lambda.$ Let $P$ be the
set $p_{1},...,p_{k}\in\partial\Lambda$ of all the entrance points
to $\Lambda$, while $Q$ be the set of the exit points
$q_{1},...,q_{k}\in
\partial\Lambda$ from $\Lambda.$ Let $\varrho_{\Lambda}=\varrho_{1}%
,...,\varrho_{k}$ be a collection of non-intersecting lattice paths
contained in $\Lambda,$ joining the points $p_{1},...,p_{k}$ and
$q_{1},...,q_{k}.$ We denote the set of such $k$-tuples of paths by
$\Omega_{\Lambda}\left( P,Q\right)  .$

We say that the measure $\mu$ is a Gibbs measure with the action
functional $S,$ if for every triple $\left(  \Lambda,P,Q\right)  $
the
conditional distributions of $\mu$ satisfy%
\begin{equation}
\mu\left(  \varrho_{\Lambda}\Bigm|\mathbf{\omega}_{\bar{\Lambda}}\right)
=\frac{\exp\left\{  S\left(  \varrho_{1}\right)  +...+S\left(  \varrho
_{k}\right)  \right\}  }{Z\left(  \Lambda,P,Q\right)  }, \label{67}%
\end{equation}
where $Z\left(  \Lambda,P,Q\right)  =\sum_{\varrho_{\Lambda}\in\Omega
_{\Lambda}\left(  P,Q\right)  }$ $\exp\left\{  S\left(  \varrho_{1}\right)
+...+S\left(  \varrho_{k}\right)  \right\}  $ is the partition function.

\subsection{\label{intermittency} The Interlacing Property}

This property of the random field $n$ holds almost surely with
respect to the measure $\mathbf{P{}r}_{K},$ as we will show below.
Its exact formulation is different at different locations and
depends on the structure of the kernel $K$ at this location. The
picture on Fig. 1 illustrates the various cases which are described
below.

\begin{figure}
[h]
\begin{center}
\includegraphics[
height=3.2727in, width=3.1208in
]%
{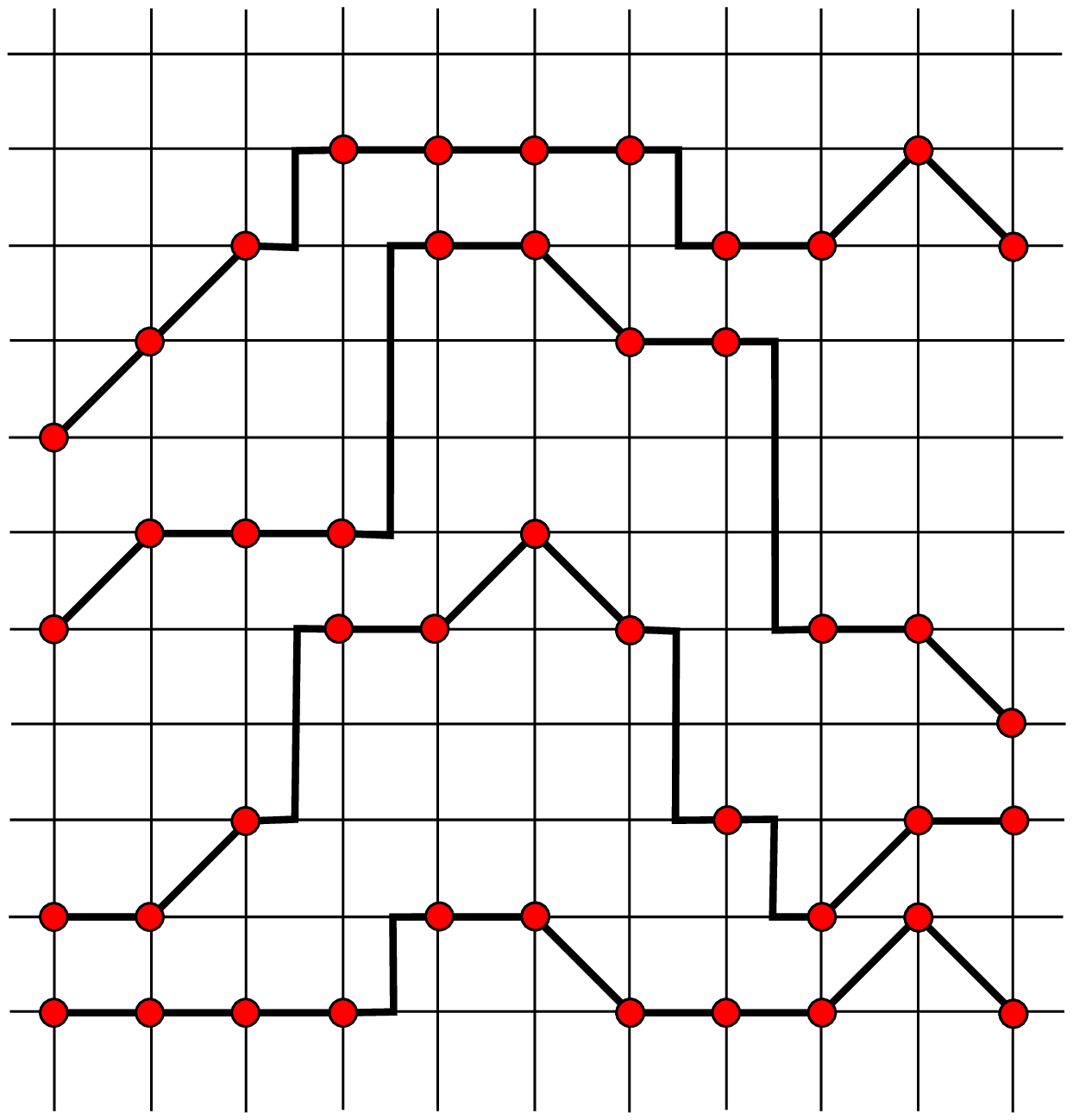}%
\caption{The first two columns, together with the 5-th and the 9-th
display ascending $\beta$-paths, the 3-rd and the 4-th -- ascending
$\alpha$-paths, the 6-th and the 10-th -- the descending
$\beta$-paths, the 7-th and the 8-th -- the descending
$\alpha$-paths.}%
\end{center}
\end{figure}

 If for some $k$ we have $\psi_{k}\left(  u\right)
=\left(  1-\alpha_{k}^{+}u\right)  ^{-1},$ then in the strip
$R_{k}=\left\{ \left(  \sigma,x\right)
\in\mathbb{Z}^{2}:\sigma=k,k+1\right\}  $ the following property
holds $\mathbf{P{}r}_{K}$-a.s.: For any two particles $n_{\left(
k,x_{1}\right)  }=n_{\left(  k,x_{2}\right)  }=1,$ $x_{1}<x_{2}$, of
the configuration $n$, separated by string of holes, $n_{\left(
k,x\right)  }=0$ for all $x_{1}<x<x_{2},$ we find on the neighboring
line $\sigma=k+1$ exactly one particle $n_{\left(  k+1,x\right) }=1$
sitting in the set $\left\{  \left(  k+1,x\right)  :x_{1}\leq
x<x_{2}\right\}  ,$ and the rest of points of this set host holes.
This is the \textit{upward} interlacing of particles.

Consider now the correspondence $\pi^{+}$, which assigns to a
particle $n_{\left( k,x\right)  }=1$ the particle $n_{\left(
k+1,\pi^{+}\left(  x\right) \right)  }=1,$ where $\pi^{+}\left(
x\right) =\min\left\{  y\geq x:n_{\left(  k+1,y\right)  }=1\right\}
\geq x.$ The correspondence $\pi^{+}$ is one-to-one with probability
one. Let us connect each particle $n_{\left(  k,x\right)  }=1$ with
the corresponding particle $n_{\left(  k+1,\pi^{+}\left(  x\right)
\right)  }=1$ by the three-link path
\begin{align}
\omega_{x,\pi^{+}\left(  x\right)  }  &  =\left[  \left(  k,x\right)
,\left( k+\tfrac{1}{2},x\right)  \right]  \cup\left[  \left(
k+\tfrac{1}{2},x\right)
,\left(  k+\tfrac{1}{2},\pi^{+}\left(  x\right)  \right)  \right] \label{66}\\
&  \cup\left[  \left(  k+\tfrac{1}{2},\pi^{+}\left(  x\right)
\right) ,\left(  k+1,\pi^{+}\left(  x\right)  \right)  \right]
.\nonumber
\end{align}
Then for different particles $n_{\left(  k,x^{\prime}\right)  }=1,$
$n_{\left(  k,x^{\prime\prime}\right)  }=1$ the connectors
$\omega_{x^{\prime },\pi^{+}\left(  x^{\prime}\right)  },$
$\omega_{x^{\prime\prime},\pi ^{+}\left(  x^{\prime\prime}\right) }$
do not intersect. We define the action $S$ on each of these
connectors by
\begin{equation}
S\left(  \omega_{x,\pi^{+}\left(  x\right)  }\right)  =\left(  \pi^{+}\left(
x\right)  -x\right)  \ln\alpha_{k}^{+}. \label{71}%
\end{equation}

In the case when $\psi_{k}\left(  u\right)  =\left(  1-\alpha_{k}^{-}%
u^{-1}\right)  ^{-1}$ the situation is very similar, except the
upward interlacing is replaced by the \textit{downward} interlacing:
The correspondence $\pi^{+}$ is replaced by $\pi^{-},$ which assigns
to a particle $n_{\left(  k,x\right)  }=1$ the particle $n_{\left(
k+1,\pi ^{-}\left(  x\right)  \right) }=1,$ where $\pi^{-}\left(
x\right) =\max\left\{  y\leq x:n_{\left( k+1,y\right) }=1\right\}
\leq x.$ Again,
$\pi^{-}$ is one-to-one a.s., and the (downward) connectors%
\begin{align}
\omega_{x,\pi^{-}\left(  x\right)  }  &  =\left[  \left(  k,x\right)
,\left( k+\tfrac{1}{2},x\right)  \right]  \cup\left[  \left(
k+\tfrac{1}{2},x\right)
,\left(  k+\tfrac{1}{2},\pi^{-}\left(  x\right)  \right)  \right] \label{65}\\
&  \cup\left[  \left(  k+\tfrac{1}{2},\pi^{-}\left(  x\right)
\right) ,\left(  k+1,\pi^{-}\left(  x\right)  \right)  \right]
\nonumber
\end{align}
do not intersect. The action $S$ is given by%

\begin{equation}
S\left(  \omega_{x,\pi^{-}\left(  x\right)  }\right)  =\left(  x-\pi
^{-}\left(  x\right)  \right)  \ln\alpha_{k}^{-}. \label{70}%
\end{equation}

For $\psi_{k}\left(  u\right)  =\left(  1+\beta_{k}^{+}u\right)  $
we have the upward interlacing of holes: If in the configuration $n$
we have two holes $n_{\left(  k,x_{1}\right)  }=n_{\left(
k,x_{2}\right)  }=0,$ $x_{1}<x_{2}$, separated by the string of
particles $\left\{  n_{\left(  k,x\right) }=1\text{ for all
}x_{1}<x<x_{2}\right\}  ,$ then on the neighboring line $\sigma=k+1$
we have $\mathbf{P{}r}_{K}$-a.s. exactly one hole $n_{\left(
k+1,x\right)  }=0$ in the set $\left\{  \left(  k+1,x\right)
:x_{1}<x\leq x_{2}\right\}  ,$ while the rest of points in this set
is filled by the particles. We then define a correspondence
$\chi^{+},$ assigning to every particle on the $\sigma=k$ line a
particle on the $\sigma=k+1$ line, as follows. Take any string of
consecutive particles $\left\{  n_{\left( k,x\right)  }=1\text{ for
all }x_{1}<x<x_{2}\right\}  $ which is maximal, i.e. $n_{\left(
k,x_{1}\right)  }=n_{\left(  k,x_{2}\right)  }=0.$ We put
$$\chi^{+}\left(  x_{1}+1\right)  =\min\left\{  x\geq
x_{1}+1:n_{\left( k+1,x\right)  }=1\right\}  ,$$ and then proceed
inductively by putting $$\chi^{+}\left(  x+1\right)  =\min\left\{
y>\chi^{+}\left(  x\right) :n_{\left(  k+1,y\right)  }=1\right\} .$$
The hole interlacing implies that $\chi^{+}$ is a.s. well-defined,
is one-to-one, and that for every particle $n_{\left( k,x\right)
}=1$ we have either $\chi^{+}\left(  x\right)  =x$ or
$\chi^{+}\left( x\right)  =x+1.$ The particle connectors, which in
this case
are segments%
\begin{equation}
\omega_{x,\chi^{+}\left(  x\right)  }=\left[  \left(  k,x\right)  ,\left(
k+1,\chi^{+}\left(  x\right)  \right)  \right]  ,\label{64}%
\end{equation}
clearly do not intersect each other. We put%
\begin{equation}
S\left(  \omega_{x,\chi^{+}\left(  x\right)  }\right)  =\left\{
\begin{array}
[c]{cc}%
0 & \text{ if }\chi^{+}\left(  x\right)  =x,\\
\ln\beta_{k}^{+} & \text{ if }\chi^{+}\left(  x\right)  =x+1.
\end{array}
\right.  \label{69}%
\end{equation}

For $\psi_{k}\left(  u\right)  =\left(  1+\beta_{k}^{-}u^{-1}\right)
$ we have likewise the downward interlacing of holes. The
correspondence $\chi^{+}$ is replaced by $\chi^{-},$ with the
property that either $\chi ^{-}\left(  x\right)  =x$ or
$\chi^{-}\left(  x\right)  =x-1.$ The connectors are
non-intersecting segments
\begin{equation}
\omega_{x,\chi^{-}\left(  x\right)  }=\left[  \left(  k,x\right)  ,\left(
k+1,\chi^{-}\left(  x\right)  \right)  \right]  , \label{63}%
\end{equation}
and we define
\begin{equation}
S\left(  \omega_{x,\chi^{-}\left(  x\right)  }\right)  =\left\{
\begin{array}
[c]{cc}%
0 & \text{ if }\chi^{-}\left(  x\right)  =x,\\
\ln\beta_{k}^{-} & \text{ if }\chi^{-}\left(  x\right)  =x-1.
\end{array}
\right.  \label{68}%
\end{equation}

\subsection{Main result}

Now we can formulate our claims more precisely. Let us fix a sequence of
functions $\psi_{k}\left(  u\right)  ,$ $k\in\mathbb{Z}^{1},$ such that for
every $k$ the function $\psi_{k}\left(  u\right)  $ is one of the four
functions $\left(  1-\alpha_{k}^{+}u\right)  ^{-1},$ $\left(  1-\alpha_{k}%
^{-}u^{-1}\right)  ^{-1},$ $\left(  1+\beta_{k}^{+}u\right)  ,$
$\left( 1+\beta_{k}^{-}u^{-1}\right)  ,$ with $\alpha_{\ast}^{\pm},$
$\beta_{\ast }^{\pm}>0$. Let us also fix a complex number $z,$ $\Im
z >0.$

\begin{theorem}\label{main}
\textbf{i) }The kernel (\ref{kernel}) defines a determinantal random
field $n$ on $\Z^2$, which is invariant with respect to the shifts
of the second coordinate.

\textbf{ii) }The random field $n$ possesses the interlacing property
as defined in Section \ref{intermittency}. In particular, there is a
map $\mathbf{\omega}$, assigning to $\mathbf{P{}r}_{K}$-a.e.
realization of $n$ a
countable collection of non-intersecting lattice paths $\mathbf{\omega}%
_{n}=\left\{  \omega_{i}\right\}  ,$ passing through all the
particles of the configuration $n.$ The construction of the
collection $\mathbf{\omega}_{n}$ is given by $\left( \ref{66}\right)
,$ $\left(  \ref{65}\right) ,$ $\left( \ref{64}\right)  $ and
$\left(  \ref{63}\right)  .$

\textbf{iii) }The random paths $\mathbf{\omega}$ thus constructed
form Gibbs Path Ensemble, as defined in Section \ref{Gibbs}. It
corresponds to the action functional $S,$ given by the formulas
$\left( \ref{71}\right)  ,$ $\left( \ref{70}\right)  ,$ $\left(
\ref{69}\right)  $ and $\left(  \ref{68}\right) .$

\textbf{iv) }For every $k\in \Z$ the matrix elements of the kernel
$K$ satisfy the following
relations:%

for the case $\psi_{k}\left(  u\right)  =\left(
1-\alpha_{k}^{+}u\right) ^{-1}$
\[
K_{k-1,\tau}\left(  x-y\right)
-\delta_{\substack{x=y\\\tau=k-1}}=K_{k,\tau }\left(  x-y\right)
-\alpha_{k}^{+}K_{k,\tau}\left(  x-y-1\right),
\]%
\[
K_{\sigma,k}\left(  x-y\right)  -\delta_{\substack{x=y\\\sigma=k}%
}=K_{\sigma,k-1}\left(  x-y\right)
-\alpha_{k}^{+}K_{\sigma,k-1}\left( x-y-1\right);
\]

for the case $\psi_{k}\left(  u\right)  =\left(
1+\beta_{k}^{+}u\right) $
\[
K_{k,\tau}\left(  x-y\right)  =\left[  K_{k-1,\tau}\left(
x-y\right) -\delta_{\substack{x=y\\\tau=k-1}}\right]
+\beta_{k}^{+}\left[  K_{k-1,\tau }\left(  x-y-1\right)
-\delta_{\substack{x=y+1\\\tau=k-1}}\right],
\]%
\[
K_{\sigma,k-1}\left(  x-y\right)  =\beta_{k}^{+}\left[
K_{\sigma,k}\left( x-y-1\right)
-\delta_{\substack{\sigma=k\\x=y+1}}\right]  +\left[
K_{\sigma,k}\left(  x-y\right)
-\delta_{\substack{\sigma=k\\x=y}}\right];
\]
and similar relations for the $\alpha^{-},\beta^{-}$ cases. The
determinant identities, expressing the properties \textbf{ii) }and
\textbf{iii)} above, are corollaries of these relations only, and
thus hold true for any other kernel $K,$ satisfying them.
\end{theorem}

\section{Examples}

\textbf{1. }Our first example will be Gibbs ensembles of the $\beta
$\textit{-paths}, introduced in $\left(  \ref{64}\right)  ,$ $\left(
\ref{63}\right)  .$ These are collections of non-intersecting
infinite paths $\left\{  \omega_{i}\right\}  $ on $\mathbb{Z}^{2},$
such that if a path visits the point $\left(  \sigma,x\right)  ,$
then its next link is either the segment $\left[  \left(
\sigma,x\right)  ,\left(  \sigma+1,x\right)  \right] $ or the
segment $\left[  \left(  \sigma,x\right)  ,\left(  \sigma
+1,x+1\right)  \right]  .$ Now let $\varrho=\varrho\left(
p,q\right)  $ be a finite piece of such $\beta$-path, where $p,q$
are the end-points of $\varrho $. We define the \textit{energy
}$U\left(  \varrho\right)  $ of this path in the following way. Let
$\varrho_{-}\left(  p,q\right)  $ be the $\beta$-path, which is the
lowest among all the $\beta$-paths connecting $p$ and $q.$ Then
$\exp\left\{  -U\left(  \varrho\right)  \right\}  $ is by definition
the area surrounded by the loop $\varrho\left(  p,q\right)
\cup\varrho_{-}\left( p,q\right)  .$ For a collection
$\mathbf{\varrho}=\left\{  \varrho _{i}\right\}  $ of finite paths
we define $H\left(  \mathbf{\varrho}\right) =\sum_{i}U\left(
\varrho_{i}\right)  .$

We call the measure $\mu$ on the ensemble $\mathbf{\omega}$ of
non-intersecting infinite $\beta$-paths the Gibbs measure
corresponding to the energy $H$ and the inverse temperature $\tau,$
if it has the following property. Let $\Lambda\subset\mathbb{Z}^{2}$
be a finite volume, and the sets $P=\left\{
p_{1},...,p_{k}\in\partial\Lambda\right\}  ,$ $Q=\left\{
q_{1},...,q_{k}\in\partial\Lambda\right\}  $ of the entrance points
and exit points are fixed. Then the conditional distribution of
$\mu$ on $\Omega _{\Lambda}\left(  P,Q\right)  =\left\{
\varrho_{\Lambda}\right\}  $ under the condition that the path
configuration $\mathbf{\omega}_{\bar{\Lambda}}$ is fixed outside
$\Lambda$ is given by the formula
\begin{equation}
\mu\left(
\varrho_{\Lambda}\Bigm|\mathbf{\omega}_{\bar{\Lambda}}\right)
=\frac{\exp\left\{  -\tau H\left(  \varrho_{\Lambda}\right)  \right\}  }%
{\bar{Z}\left(  \Lambda,P,Q\right)  }.\label{49}%
\end{equation}

This definition is just a convenient rewriting of the one given
above. The advantage is that our function $H$ here is manifestly
translation-invariant. Our main result implies in particular that
the determinantal random fields $\mu_{\varkappa,z}$ defined by the
kernel $K=K\left(  \varkappa,z\right)  $ with the functions
$\psi_{k}\left(  u\right)  =\left(  1+\varkappa e^{k\tau }u\right)
$, interpreted as path measures, are Gibbs measures with the energy
$H$ and the inverse temperature $\tau.$ Here $\varkappa>0$ is any
real number.

When the temperature $\tau^{-1}$ goes to zero, the Gibbs measures
$\left( \ref{49}\right)  $ are concentrated on ground-state
configurations, which are local minima of the energy $H.$ For low
temperatures they are concentrated on configurations which are small
perturbations of the ground state configurations, see Fig. 2. Note
that the ground state configurations have their corner points
confined to at most two nearest neighbor vertical lattice lines. One
can say that for large $\tau$ our two-dimensional random field is
essentially one-dimensional, and outside the strip of width
$\sim\tau^{-1}$ it is basically frozen. Along this vertical
direction it has the following correlation decay property: for every
two
finite subsets $A,B\subset\mathbb{Z}^{2}$ we have $\left\langle n_{A+x}%
n_{B}\right\rangle -\left\langle n_{A}\right\rangle \left\langle
n_{B}\right\rangle \rightarrow0$ as $\left\vert x\right\vert
\rightarrow \infty.$ Here $n_{A}=\prod_{\left(  \sigma,x\right)  \in
A}n_{\sigma,x}.$

\begin{figure}
[h]
\begin{center}
\includegraphics[
height=3.9452in, width=2.3462in
]%
{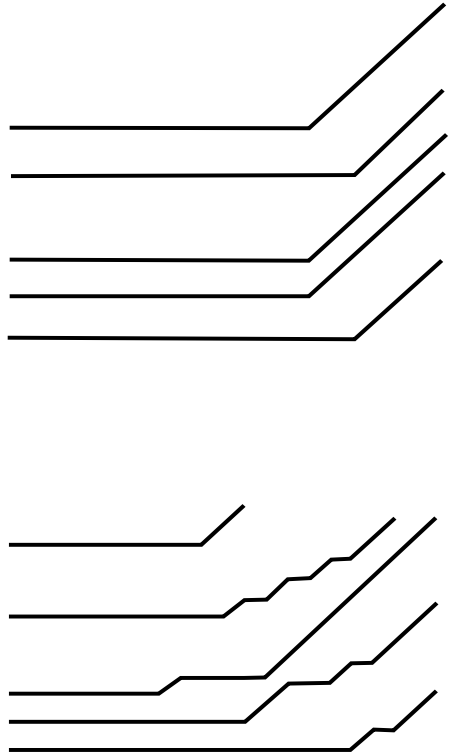}%
\caption{A ground state configuration of the $\beta$-paths, and a
low-temperature configuration.}
\end{center}
\end{figure}

Without loss of generality we can assume that $\left\vert
z\right\vert =1.$
The parameter $z$ then defines the \textquotedblleft slope\textquotedblright%
\ of the \textquotedblleft height function\textquotedblright, or,
what is the same, the (constant) density of the paths in the path
ensemble.

Note now, that for every $z$ the determinantal processes
$\mu_{\varkappa,z}$ are different for different values of
$\varkappa.$ This follows just from the computation of the second
correlation function for these processes. That means that there are
continuum non-translation-invariant Gibbs measures $\left(
\ref{49}\right)  ,$ corresponding to the same slope and the same
temperature. Of course, the field $\mu_{\varkappa,z}$ is just a
translate of the field $\mu_{\tau\varkappa,z}$ by the unit lattice
vector. But the fields $\mu_{\varkappa,z}$ with $\varkappa$ between
$1$ and $\tau$ are all different and are not related by the lattice
shift transformation.

One can understand better the role played by the parameter
$\varkappa$ looking at the boundary conditions and the limiting
behavior of the processes $\mu_{\varkappa,z}.$ The restriction of
the process to any vertical line $\sigma=const$ gives the
(translation-invariant) sine process with density $\arg z/\pi$. Let
us consider two such lines, say $\sigma=\pm k.$ Then the paths of
the process $\mu_{\varkappa,z}$ define in a natural way the coupling
$\mathcal{C}\left(  k,\varkappa,\tau\right) $ between the two sine
processes. Consider the limiting coupling $\mathcal{C}\left(
\varkappa,\tau\right)  =\lim_{k\rightarrow\infty }\mathcal{C}\left(
k,\varkappa,\tau\right)  .$ It turns out that the couplings
$\mathcal{C}\left(  \varkappa,\tau\right)  $ are still non-trivial
(i.e. these couplings are not product-couplings), and are different
for different $\varkappa$. The two couplings $\mathcal{C}\left(
\varkappa ,\tau\right)  $ and $\mathcal{C}\left(
\tau\varkappa,\tau\right)  $ are related by the unit shift of one of
the $\sin$-processes. The couplings $\mathcal{C}\left(
\varkappa,\tau\right)  $ become trivial only in the limit when
$\tau\rightarrow0.$ In this limit the fields $\mu_{\varkappa,z}$
become fully translation invariant.

A straightforward geometric interpretation of our ensemble of the
$\beta $-paths is to relate them to the lozenge tilings of the
plane. Our paths are then composed by the middle lines of all
\textquotedblleft vertical\textquotedblright\ plaquettes, see Fig.
3, cf. \cite{J2}.

\begin{figure}
[ptb]
\begin{center}
\includegraphics[
height=3.0676in, width=2.6642in
]%
{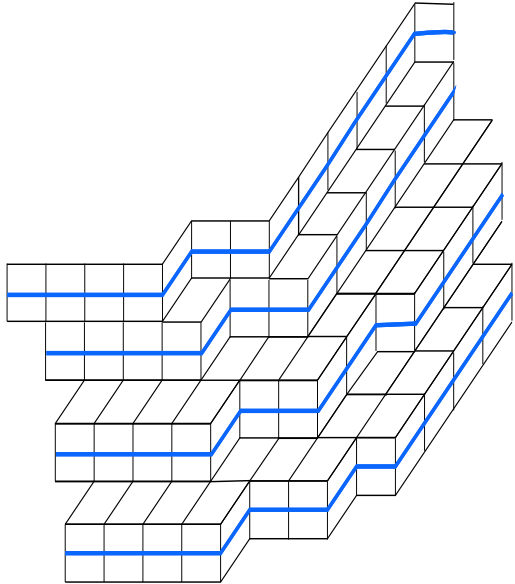}%
\caption{The $\beta$-paths and the corresponding (deformed-)lozenge tiling.}%
\end{center}
\end{figure}

One may wonder about the existence of the asymptotic shape of the
height function corresponding to the field $\mu_{1,z}$, i.e. to the
choice $\psi _{k}\left(  u\right)  =\left(  1+\tau^{k}u\right)  .$
However, this height function is almost frozen outside the strip of
width $\sim\tau^{-1}$ around the line $\sigma=0.$ If we scale this
surface by the factor $\tau$ in all three dimensions, then the
conjectural limit when $\tau\rightarrow0$ is a non-random
cylindrical surface. This surface has a gutter shape, see Fig. 4,
and is given by the following geometric construction.

\begin{figure}
[ptb]
\begin{center}
\includegraphics[
height=2.6592in, width=4.005in
]%
{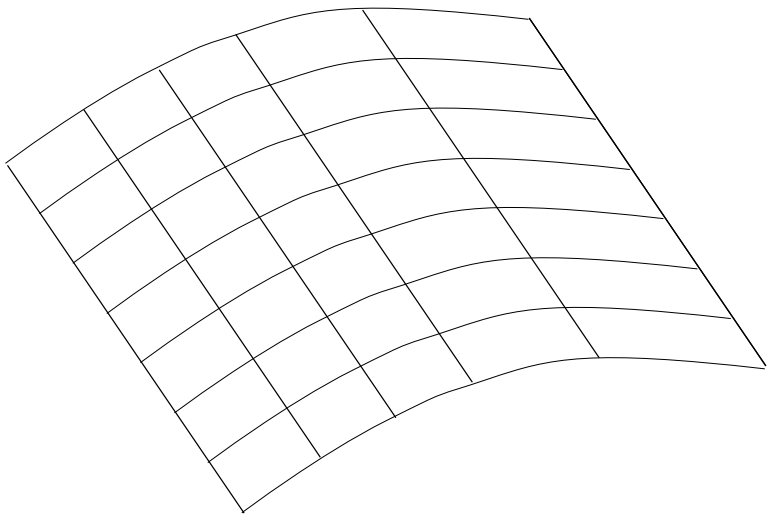}%
\caption{The limit shape}%
\end{center}
\end{figure}

To describe it we first recall the geometric construction (see
\cite{S1} or \cite{S2}), used to obtain the limit shape
$\mathcal{C}_{CK}$ of the plane partitions by Cerf and Kenyon in
\cite{CK}. For every positive unit vector
$\mathbf{n}\in\Delta^{2}=$ $S^{2}\cap\mathbb{R}_{+}^{3}$ let $\mathrm{ent}%
\left(  \mathbf{n}\right)  $ be the residual entropy of the lozenge
tilings having the slope plane orthogonal to $\mathbf{n}$ (see
Theorem 1.1 in \cite{CK}). Now define the halfspaces
\[
K_{\mathbf{n}}=\left\{  \mathbf{x}\in\mathbb{R}^{3}:\;\left(  \mathbf{x}%
,\mathbf{n}\right)  \geq\,\mathrm{ent}\left(  \mathbf{n}\right)
\right\}  .
\]
Let
\[
K=\cap_{\mathbf{n}\in\Delta^{2}}K_{\mathbf{n}};
\]
the boundary of the region $K$ is precisely the surface
$\mathcal{C}_{CK},$ describing the typical shape of a large plane
partition.

To describe the gutter shape we first define its slope, $\gamma$.
This is determined by the frequency of the lines in our family
(indeed, they are just the level lines of the height function). If
$z=e^{i\varphi},$ then the density in question is the first
correlation function of our determinantal process, and it is equal
to $\frac{\varphi}{\pi}.$ Therefore $\gamma$ satisfies
$\tan\gamma=\frac{\varphi}{1-\varphi}.$ Let us define the vector
$\mathbf{m}\left(  \gamma\right)  =\left(  m_{x},m_{y},m_{z}\right)
=\left( -\frac{1}{\sqrt{2}},-\frac{1}{\sqrt{2}},\tan\gamma\right)
.$ This is the direction of our gutter. Consider the arc $A\left(
\gamma\right)  $ of vectors in the \textquotedblleft
triangle\textquotedblright\ $\Delta^{2},$
which are orthogonal to $\mathbf{m}\left(  \gamma\right)  :$%
\[
A\left(  \gamma\right)  =\left\{  \mathbf{n}\in\Delta^{2}:\left(
\mathbf{n},\mathbf{m}\left(  \gamma\right)  \right)  =0\right\}  .
\]
Our gutter surface, $G\left(  \gamma\right)  ,$ is defined to be the
boundary of the convex region
\[
K\left(  \gamma\right)  \equiv\cap_{\mathbf{n}\in A\left(
\gamma\right) }K_{\mathbf{n}}.
\]
Note that the surface $G\left(  \gamma\right)  $ consists of
straight lines parallel to the vector $\mathbf{m}\left(
\gamma\right)  .$ The surfaces $G\left(  \gamma\right)  $ and
$\mathcal{C}_{CK}$ are tangent to each other along the common curve
$g\left(  \gamma\right) =G\left(  \gamma\right)
\cap\mathcal{C}_{CK}.$ Each of the curves $g\left(  \gamma\right)  $
is a smooth curve without straight pieces. Asymptotically each of
them approaches
the Vershik curve $\mathcal{C}_{V}:\left\{  \exp\left(  -\tfrac{\pi}{\sqrt{6}%
}x\right)  +\exp\left(  -\tfrac{\pi}{\sqrt{6}}y\right)
=1,z=0\right\}  ,$
which belongs to the boundary of the curved part of the surface $\mathcal{C}%
_{CK}.$

\textbf{2. }Our second example is the ensemble of
$\alpha\beta$-paths, with $\psi_{2k}\left(  u\right)  =\left(
1-\alpha_{k}^{-}u^{-1}\right)  ^{-1}$ and $\psi_{2k+1}\left(
u\right)  =\left(  1+\beta_{k}^{+}u\right)  .$ Again we will choose
$\alpha$ and $\beta$ to be geometric progressions, by putting
$\alpha_{k}^{-}=(\varkappa e^{k\tau})^{-1},$ $\beta_{k}^{+}=\lambda
e^{k\tau},$ with $\varkappa,\lambda,\tau>0.$ In the same way that
the $\beta$-paths are related to the lozenge tiling, the
$\alpha\beta$-paths are related to the domino tilings. The relation
however is not so easy to explain; the corresponding construction,
establishing the bijection between the two entities, is presented in
\cite{J}, see also \cite{LRS}.

The Fig. 5 shows one collection of $\alpha\beta$-paths with
$\varkappa=\lambda=1$ and $\tau=\infty$ (which therefore should be
called a ground state configuration).

\begin{figure}
[ptb]
\begin{center}
\includegraphics[
height=4.9913in, width=3.2885in
]%
{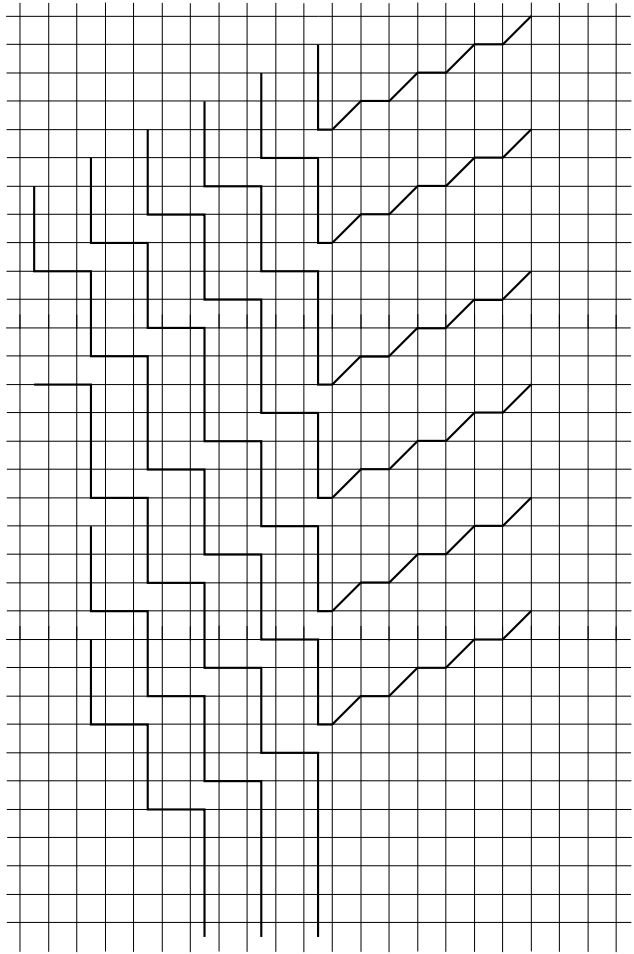}%
\caption{A ground state configuration of $\alpha\beta$-paths.}%
\end{center}
\end{figure}

For $k>0$ all the $\alpha$-steps are zero height steps, while all
the $\beta$-steps are ascending. For $k<0$ all the $\beta$-steps are
zero height steps, while all the $\alpha$-steps are descending in a
maximal possible way. The Fig. 6 shows the corresponding domino
tiling.

\begin{figure}
[ptb]
\begin{center}
\includegraphics[
height=3.2719in, width=3.3192in
]%
{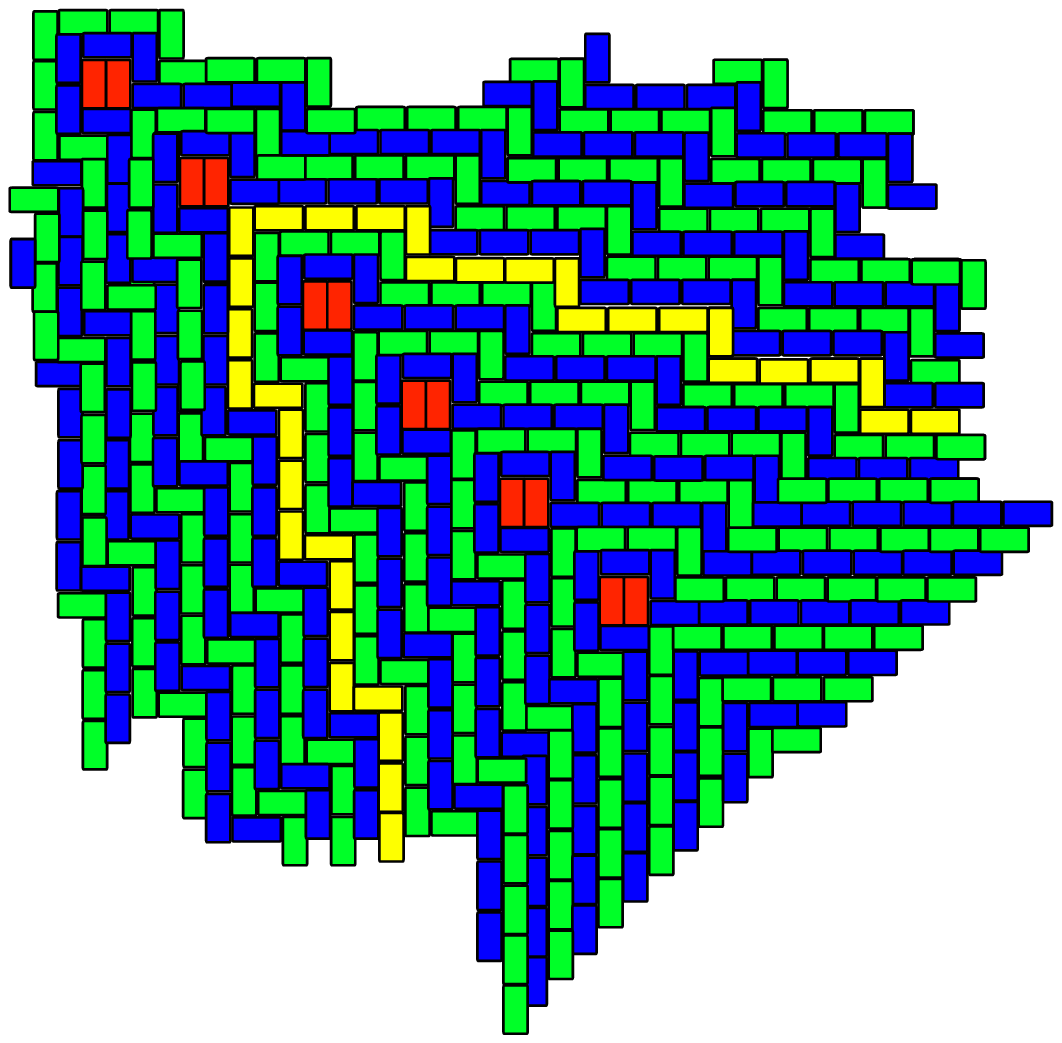}%
\caption{The domino tiling, corresponding to the paths above.}%
\end{center}
\end{figure}

In domino tilings, the elementary moves correspond to finding a
$2\times2$ square tiled by two dominoes, say horizontal, and
replacing this pair of dominos by two vertical ones. If one assigns
four weights $a,b,c,d$ to the four possible positions of a single
domino, then every move replacing a horizontal pair by a vertical
one changes the overall weight by a constant factor. (If $a,b$ are
two horizontal weights, then the overall change will be by a factor
$\frac{cd}{ab}.$) In our case the overall weight after an elementary
move depends on the parity of the $2\times2$ square, and is
$\frac{\varkappa}{\lambda}$ in one case, and $\frac{\varkappa}{\lambda}%
e^{\tau}$ in the other. This means that the measures on the domino
tilings that we have constructed, can not be obtained by assigning
weights to single dominoes.

Again, for $\tau\neq0$ our measures are non-translation-invariant,
and by varying the ratio $\frac{\varkappa}{\lambda}$ we obtain a
whole continuum of different measures.

In the case $\tau=0$, the two-parametric measure on lozenge tilings
and the three-parametric measure on domino tilings are translation
invariant with respect to all shifts of $\Z^2$. These measures are
well known; for lozenge tilings they were obtained in \cite{Ken},
\cite{OR}, and for domino tilings they were obtained in \cite{BP},
\cite{CKP}, see also \cite{J}. As proved in \cite{Sheffield}, they
are the only fully translation invariant ergodic measures.

\section{Proof of the Main Result}

The proof of \textbf{i} of Theorem \ref{main} is given in Section
\ref{positivity} below.

We will start the proof of \textbf{ii}-\textbf{iv} by dealing with
the special case when for each $k\in\mathbb{Z}^{1}$ the function
$\psi_{k}\left( u\right) $ is either $\left(  1-\alpha_{k}u\right)
^{-1}$ or $\left( 1+\beta_{k}u\right)  ,$ with
$\alpha_{k},\beta_{k}$ some positive constants. We will consider the
general case at the end of the proof.

\subsection{Linear relations}

Let the sequence of functions $\psi_{k}\left(  u\right)  $ be given, where for
every $k\in\mathbb{Z}^{1}$ the function $\psi_{k}\left(  u\right)  $ is either
$\left(  1-\alpha_{k}u\right)  ^{-1}$ or $\left(  1+\beta_{k}u\right)  ,$ with
$\alpha_{k},\beta_{k}$ some positive constants.

In this subsection we will show that the kernel
$K_{\sigma,\tau}\left( x-y\right)  $ satisfies the linear relations
mentioned in Theorem \ref{main}. Indeed, if $\psi_{k}\left(
u\right) =\left( 1-\alpha_{k}u\right)  ^{-1},$ then for $\sigma=k-1$
and $\tau\geq k$
\begin{align*}
K_{k-1,\tau}\left(  x-y\right)   &  =\frac{1}{2\pi
i}\int_{C_+}\left( 1-\alpha_{k}u\right)  \left(
\prod_{k+1}^{\tau}\psi_{j}\left( u\right)
\right)  ^{-1}\frac{du}{u^{x-y+1}}\\
&  =K_{k,\tau}\left(  x-y\right)  -\alpha_{k}K_{k,\tau}\left(  x-y-1\right)  ,
\end{align*}
the same for $\tau<k-1:$%
\begin{align*}
&  K_{k,\tau}\left(  x-y\right)  -\alpha_{k}K_{k,\tau}\left(  x-y-1\right)  \\
&  =\frac{1}{2\pi i}\int_{C_-}\prod_{\tau+1}^{k-1}\psi_{j}\left(
u\right) \left(  1-\alpha_{k}u\right)
^{-1}\frac{du}{u^{x-y+1}}-\alpha _{k}\frac{1}{2\pi
i}\int_{C_-}\prod_{\tau+1}^{k-1}\psi_{j}\left(
u\right)  \left(  1-\alpha_{k}u\right)  ^{-1}\frac{du}{u^{x-y}}\\
&  =\frac{1}{2\pi i}\int_{C_-}\prod_{\tau+1}^{k-1}\psi_{j}\left(
u\right) \frac{du}{u^{x-y+1}}=K_{k-1,\tau}\left(  x-y\right)
\end{align*}
while for $\tau=k-1$ we have
\begin{align*}
K_{k-1,k-1}\left(  x-y\right)   &  =\frac{1}{2\pi i}\int_{C_+}%
\frac{du}{u^{x-y+1}},\\
  K_{k,k-1}\left(  x-y\right)  -&\alpha_{k}K_{k,k-1}\left(  x-y-1\right)  \\
&  =\frac{1}{2\pi i}\int_{C_-}\left(  1-\alpha_{k}u\right)
^{-1}\frac{du}{u^{x-y+1}}-\alpha_{k}\frac{1}{2\pi i}\int_{C_-}\left(
1-\alpha_{k}u\right)  ^{-1}\frac{du}{u^{x-y}}\\
&  =\frac{1}{2\pi i}\int_{C_-}\frac{du}{u^{x-y+1}},
\end{align*}
which means that
\begin{equation}
K_{k-1,k-1}\left(  x-y\right)  -K_{k,k-1}\left(  x-y\right)  +\alpha
_{k}K_{k,k-1}\left(  x-y-1\right)  =\delta_{x=y}.\label{83}%
\end{equation}
Altogether, these relations read%
\begin{equation}
K_{k-1,\tau}\left(  x-y\right)  -\delta_{\substack{x=y\\\tau=k-1}}=K_{k,\tau
}\left(  x-y\right)  -\alpha_{k}K_{k,\tau}\left(  x-y-1\right)  .\label{90}%
\end{equation}

Also, if $\tau=k$ and $\sigma\leq k-1,$ then%
\begin{align*}
K_{\sigma,k}\left(  x-y\right)   &  =\frac{1}{2\pi
i}\int_{C_+}\left( \prod_{\sigma+1}^{\tau}\psi_{j}\left(  u\right)
\right) ^{-1}\left(
1-\alpha_{k}u\right)  \frac{du}{u^{x-y+1}}\\
&  =K_{\sigma,k-1}\left(  x-y\right)  -\alpha_{k}K_{\sigma,k-1}\left(
x-y-1\right)  ,
\end{align*}
and the same for $\sigma>k.$ Since the diagonal elements $K_{r,r}\left(
x-y\right)  $ do not depend on $r,$ for $\sigma=k$ we have immediately from
$\left(  \ref{83}\right)  :$
\[
K_{k,k}\left(  x-y\right)  -K_{k,k-1}\left(  x-y\right)  +\alpha_{k}%
K_{k,k-1}\left(  x-y-1\right)  =\delta_{x=y}.
\]
Altogether,%
\begin{equation}
K_{\sigma,k}\left(  x-y\right)  -\delta_{\substack{x=y\\\sigma=k}%
}=K_{\sigma,k-1}\left(  x-y\right)  -\alpha_{k}K_{\sigma,k-1}\left(
x-y-1\right)  . \label{88}%
\end{equation}

Likewise, for $\psi_{k}\left(  u\right)  =\left(  1+\beta_{k}u\right)  $,
$\sigma=k$ and $k-1>\tau$ we have
\begin{align*}
K_{k,\tau}\left(  x-y\right)   &  =\frac{1}{2\pi i}\int_{C_-}\left(
\prod_{\tau+1}^{k-1}\psi_{j}\left( u\right)  \right)  \left( 1+\beta
_{k}u\right)  \frac{du}{u^{x-y+1}}\\
&  =K_{k-1,\tau}\left(  x-y\right)  +\beta_{k}K_{k-1,\tau}\left(
x-y-1\right)  ,
\end{align*}
and similarly for $k-1<\tau.$ For $\tau=k-1$ we have
\begin{equation}
K_{k,k-1}\left(  x-y\right)  =\frac{1}{2\pi i}\int_{C_-}\left(
1+\beta_{k}u\right)  \frac{du}{u^{x-y+1}}, \label{86}%
\end{equation}
while%
\begin{equation}
K_{k-1,k-1}\left(  x-y\right)  +\beta_{k}K_{k-1,k-1}\left(
x-y-1\right) =\frac{1}{2\pi
i}\int_{C_+}\frac{du}{u^{x-y+1}}+\beta_{k}\frac{1}{2\pi
i}\int_{C_+}\frac{du}{u^{x-y}}, \label{85}%
\end{equation}
so%
\begin{align}
&  K_{k-1,k-1}\left(  x-y\right)  +\beta_{k}K_{k-1,k-1}\left(  x-y-1\right)
-K_{k,k-1}\left(  x-y\right) \label{84}\\
&  =\frac{1}{2\pi i}%
%TCIMACRO{\doint }%
%BeginExpansion
{\displaystyle\oint}
%EndExpansion
\left(  1+\beta_{k}u\right)  \frac{du}{u^{x-y+1}}=\delta_{x=y}+\beta_{k}%
\delta_{x=y+1}.
\end{align}
Summarizing, we have%
\begin{equation}
K_{k,\tau}\left(  x-y\right)  =\left[  K_{k-1,\tau}\left(  x-y\right)
-\delta_{\substack{x=y\\\tau=k-1}}\right]  +\beta_{k}\left[  K_{k-1,\tau
}\left(  x-y-1\right)  -\delta_{\substack{x=y+1\\\tau=k-1}}\right]  .
\label{82}%
\end{equation}

The last relation we obtain by considering for $\psi_{k}\left(  u\right)
=\left(  1+\beta_{k}u\right)  $ the case when $\sigma>k$ while $\tau=k-1.$
Then we have%
\begin{align*}
K_{\sigma,k-1}\left(  x-y\right)   &  =\frac{1}{2\pi
i}\int_{C_-}\left(  1+\beta_{k}u\right)  \left(
\prod_{k+1}^{\sigma}\psi_{j}\left(
u\right)  \right)  \frac{du}{u^{x-y+1}}\\
&  =K_{\sigma,k}\left(  x-y\right)  +\beta_{k}K_{\sigma,k}\left(
x-y-1\right)  .
\end{align*}
The same relation holds in the region $\sigma<k,$ while for $\sigma=k$ we use
$\left(  \ref{84}\right)  $, which immediately imply that
\[
K_{k,k}\left(  x-y\right)  +\beta_{k}K_{k,k}\left(  x-y-1\right)
-K_{k,k-1}\left(  x-y\right)  =\delta_{x=y}+\beta_{k}\delta_{x=y+1},
\]
thus getting us to%
\begin{equation}
K_{\sigma,k-1}\left(  x-y\right)  =\beta_{k}\left[  K_{\sigma,k}\left(
x-y-1\right)  -\delta_{\substack{\sigma=k\\x=y+1}}\right]  +\left[
K_{\sigma,k}\left(  x-y\right)  -\delta_{\substack{\sigma=k\\x=y}}\right]  .
\label{81}%
\end{equation}

\subsection{\label{support} Interlacing property. Simplest case.}

Let us start by checking the interlacing property in the simplest
situations. In the case $\psi_{k}\left(  u\right)  =\left(  1-\alpha
_{k}u\right)  ^{-1}$ we will show that
\begin{equation}
\mathbf{P{}r}_{K}\left\{  \left(
\begin{array}
[c]{cc}%
1 & \ast\\
1 & 0
\end{array}
\right)  _{k-1,k}\right\}  =0,~\ \mathbf{P{}r}_{K}\left\{  \left(
\begin{array}
[c]{cc}%
0 & 1\\
\ast & 1
\end{array}
\right)  _{k-1,k}\right\}  =0, \label{87}%
\end{equation}
where the symbol $\left(
\begin{array}
[c]{cc}%
1 & \ast\\
1 & 0
\end{array}
\right)  _{k-1,k}$ denotes the corresponding event in some two by
two square in the vertical strip $\left\{  \left(  k-1,\ast\right)
,\left( k,\ast\right) \right\}  .$ For the case $\sigma=k$ with
$\psi_{k}\left(  u\right)  =\left(
1+\beta_{k}u\right)  $ we will show that%
\[
\mathbf{P{}r}_{K}\left\{  \left(
\begin{array}
[c]{cc}%
0 & 1\\
0 & \ast
\end{array}
\right)  _{k-1,k}\right\}  =0,~\ \mathbf{P{}r}_{K}\left\{  \left(
\begin{array}
[c]{cc}%
\ast & 0\\
1 & 0
\end{array}
\right)  _{k-1,k}\right\}  =0.
\]
To save on notation, we will put $k=1,$ and we will write $\alpha,\beta,\psi$
instead of $\alpha_{1},\beta_{1},\psi_{1}.$

\textbf{1. }The case $\psi\left(  u\right)  =\left(  1-\alpha u\right)
^{-1}:$
\[
\mathbf{P{}r}_{K}\left\{  \left(
\begin{array}
[c]{cc}%
1 & \ast\\
1 & 0
\end{array}
\right)  _{0,1}\right\}  =0.
\]
This relation is equivalent to showing that
\[
\det\left\vert
\begin{array}
[c]{ccc}%
K_{0,0}\left(  0\right)  & K_{0,0}\left(  -1\right)  & K_{0,1}\left(  0\right)
\\
K_{0,0}\left(  1\right)  & K_{0,0}\left(  0\right)  & K_{0,1}\left(  1\right)
\\
K_{1,0}\left(  0\right)  & K_{1,0}\left(  -1\right)  & K_{1,1}\left(
0\right)  -1
\end{array}
\right\vert =0.
\]
But this relation does hold, since the relation $\left(  \ref{88}\right)  $
implies that the last column is a linear combination of the remaining two.

\textbf{2. }The case $\psi\left(  u\right)  =\left(  1-\alpha u\right)
^{-1}:$
\[
\mathbf{P{}r}_{K}\left\{  \left(
\begin{array}
[c]{cc}%
0 & 1\\
\ast & 1
\end{array}
\right)  _{0,1}\right\}  =0.
\]
We have to check that
\[
\det\left\vert
\begin{array}
[c]{ccc}%
K_{1,1}\left(  0\right)  & K_{1,1}\left(  -1\right)  & K_{1,0}\left(
-1\right) \\
K_{1,1}\left(  1\right)  & K_{1,1}\left(  0\right)  & K_{1,0}\left(  0\right)
\\
K_{0,1}\left(  1\right)  & K_{0,1}\left(  0\right)  & K_{0,0}\left(  0\right)
-1
\end{array}
\right\vert =0.
\]
But from $\left(  \ref{90}\right)  $ it follows that the last row is a
combination of the remaining two.

\textbf{3.} The case $\psi\left(  u\right)  =\left(  1+\beta u\right)  :$%
\begin{equation}
\mathbf{P{}r}_{K}\left\{  \left(
\begin{array}
[c]{cc}%
0 & 1\\
0 & \ast
\end{array}
\right)  _{0,1}\right\}  =0.~\ \label{75}%
\end{equation}
We have thus to show the vanishing of
\[
\det\left\vert
\begin{array}
[c]{ccc}%
K_{0,0}\left(  0\right)  -1 & K_{0,0}\left(  -1\right)  & K_{0,1}\left(
-1\right) \\
K_{0,0}\left(  1\right)  & K_{0,0}\left(  0\right)  -1 & K_{0,1}\left(
0\right) \\
K_{1,0}\left(  1\right)  & K_{1,0}\left(  0\right)  & K_{1,1}\left(  0\right)
\end{array}
\right\vert .
\]
But the relation $\left(  \ref{82}\right)  $ tells us that the third row of
the last determinant is a linear combination of the first two.

\textbf{4. }The case $\psi\left(  u\right)  =\left(  1+\beta u\right)  :$%
\[
\mathbf{P{}r}_{K}\left\{  \left(
\begin{array}
[c]{cc}%
\ast & 0\\
1 & 0
\end{array}
\right)  _{0,1}\right\}  =0.
\]
We thus need the vanishing of the determinant%
\[
\det\left\vert
\begin{array}
[c]{ccc}%
K_{0,0}\left(  0\right)  & K_{0,1}\left(  0\right)  & K_{0,1}\left(  -1\right)
\\
K_{1,0}\left(  0\right)  & K_{1,1}\left(  0\right)  -1 & K_{1,1}\left(
-1\right) \\
K_{1,0}\left(  1\right)  & K_{1,1}\left(  1\right)  & K_{1,1}\left(  0\right)
-1
\end{array}
\right\vert .
\]
But the first column is a combination of the second and the third, due to
$\left(  \ref{81}\right)  .$

\subsection{\label{emoves} Elementary moves.}

Here we will prove another set of identities, corresponding to the elementary
moves of the paths. Since every move involves two adjacent columns of the
lattice, we have four different types of moves, according to the four types --
$\alpha\alpha,$ $\alpha\beta,$ $\beta\alpha,$ or $\beta\beta$ -- of the
columns pair.

%\begin{remark}
%The case of the lozenge tilings corresponds to the $\beta\beta$ case. The
%elementary move there corresponds to the replacement of a tiling of a hexagon
%by three different lozenges by another one by the same three lozenges.
%\end{remark}

\textbf{1. }We start with the case $\psi_{1}\left(  u\right)  =\left(
1+\beta_{1}u\right)  ,$ $\psi_{2}\left(  u\right)  =\left(  1+\beta
_{2}u\right)  .$ We will prove that
\[
\beta_{1}\mathbf{P{}r}_{K}\left\{  \left(
\begin{array}
[c]{ccc}%
\ast & 0 & 1\\
1 & 1 & \ast
\end{array}
\right)  _{0,1,2}\right\}  =\beta_{2}\mathbf{P{}r}_{K}\left\{  \left(
\begin{array}
[c]{ccc}%
\ast & 1 & 1\\
1 & 0 & \ast
\end{array}
\right)  _{0,1,2}\right\}  .
\]
The corresponding determinant relation reads:%
\begin{align}
&  \beta_{1}\det\left\vert
\begin{array}
[c]{cccc}%
K_{0,0}\left(  0\right)  & K_{0,1}\left(  0\right)  & K_{0,1}\left(  -1\right)
& K_{0,2}\left(  -1\right) \\
K_{1,0}\left(  0\right)  & K_{1,1}\left(  0\right)  & K_{1,1}\left(  -1\right)
& K_{1,2}\left(  -1\right) \\
K_{1,0}\left(  1\right)  & K_{1,1}\left(  1\right)  & K_{1,1}\left(  0\right)
-1 & K_{1,2}\left(  0\right) \\
K_{2,0}\left(  1\right)  & K_{2,1}\left(  1\right)  & K_{2,1}\left(  0\right)
& K_{2,2}\left(  0\right)
\end{array}
\right\vert \label{61}\\
&  =\beta_{2}\det\left\vert
\begin{array}
[c]{cccc}%
K_{0,0}\left(  0\right)  & K_{0,1}\left(  0\right)  & K_{0,1}\left(  -1\right)
& K_{0,2}\left(  -1\right) \\
K_{1,0}\left(  0\right)  & K_{1,1}\left(  0\right)  -1 & K_{1,1}\left(
-1\right)  & K_{1,2}\left(  -1\right) \\
K_{1,0}\left(  1\right)  & K_{1,1}\left(  1\right)  & K_{1,1}\left(  0\right)
& K_{1,2}\left(  0\right) \\
K_{2,0}\left(  1\right)  & K_{2,1}\left(  1\right)  & K_{2,1}\left(  0\right)
& K_{2,2}\left(  0\right)
\end{array}
\right\vert .\nonumber
\end{align}

Due to the relation $\left(  \ref{81}\right)  ,$ applied to the first
determinant,
\[
\det\left\vert
\begin{array}
[c]{cccc}%
K_{0,0}\left(  0\right)  & K_{0,1}\left(  0\right)  & \beta_{1}K_{0,1}\left(
-1\right)  & K_{0,2}\left(  -1\right) \\
K_{1,0}\left(  0\right)  & K_{1,1}\left(  0\right)  & \beta_{1}K_{1,1}\left(
-1\right)  & K_{1,2}\left(  -1\right) \\
K_{1,0}\left(  1\right)  & K_{1,1}\left(  1\right)  & \beta_{1}\left[
K_{1,1}\left(  0\right)  -1\right]  & K_{1,2}\left(  0\right) \\
K_{2,0}\left(  1\right)  & K_{2,1}\left(  1\right)  & \beta_{1}K_{2,1}\left(
0\right)  & K_{2,2}\left(  0\right)
\end{array}
\right\vert ,
\]
subtraction from the third column the first one and adding the second one,
results in%
\[
\det\left\vert
\begin{array}
[c]{cccc}%
K_{0,0}\left(  0\right)  & K_{0,1}\left(  0\right)  & 0 & K_{0,2}\left(
-1\right) \\
K_{1,0}\left(  0\right)  & K_{1,1}\left(  0\right)  & 1 & K_{1,2}\left(
-1\right) \\
K_{1,0}\left(  1\right)  & K_{1,1}\left(  1\right)  & 0 & K_{1,2}\left(
0\right) \\
K_{2,0}\left(  1\right)  & K_{2,1}\left(  1\right)  & 0 & K_{2,2}\left(
0\right)
\end{array}
\right\vert =-\det\left\vert
\begin{array}
[c]{ccc}%
K_{0,0}\left(  0\right)  & K_{0,1}\left(  0\right)  & K_{0,2}\left(  -1\right)
\\
K_{1,0}\left(  1\right)  & K_{1,1}\left(  1\right)  & K_{1,2}\left(  0\right)
\\
K_{2,0}\left(  1\right)  & K_{2,1}\left(  1\right)  & K_{2,2}\left(  0\right)
\end{array}
\right\vert .
\]
Due to the relation $\left(  \ref{82}\right)  ,$ applied to the second
determinant,
\[
\det\left\vert
\begin{array}
[c]{cccc}%
K_{0,0}\left(  0\right)  & K_{0,1}\left(  0\right)  & K_{0,1}\left(  -1\right)
& K_{0,2}\left(  -1\right) \\
\beta_{2}K_{1,0}\left(  0\right)  & \beta_{2}\left[  K_{1,1}\left(  0\right)
-1\right]  & \beta_{2}K_{1,1}\left(  -1\right)  & \beta_{2}K_{1,2}\left(
-1\right) \\
K_{1,0}\left(  1\right)  & K_{1,1}\left(  1\right)  & K_{1,1}\left(  0\right)
& K_{1,2}\left(  0\right) \\
K_{2,0}\left(  1\right)  & K_{2,1}\left(  1\right)  & K_{2,1}\left(  0\right)
& K_{2,2}\left(  0\right)
\end{array}
\right\vert ,
\]
subtraction from the second row the last one and adding the third one, results
in:%
\begin{align*}
&  \det\left\vert
\begin{array}
[c]{cccc}%
K_{0,0}\left(  0\right)  & K_{0,1}\left(  0\right)  & K_{0,1}\left(  -1\right)
& K_{0,2}\left(  -1\right) \\
0 & 0 & 1 & 0\\
K_{1,0}\left(  1\right)  & K_{1,1}\left(  1\right)  & K_{1,1}\left(  0\right)
& K_{1,2}\left(  0\right) \\
K_{2,0}\left(  1\right)  & K_{2,1}\left(  1\right)  & K_{2,1}\left(  0\right)
& K_{2,2}\left(  0\right)
\end{array}
\right\vert \\
&  =-\det\left\vert
\begin{array}
[c]{ccc}%
K_{0,0}\left(  0\right)  & K_{0,1}\left(  0\right)  & K_{0,2}\left(  -1\right)
\\
K_{1,0}\left(  1\right)  & K_{1,1}\left(  1\right)  & K_{1,2}\left(  0\right)
\\
K_{2,0}\left(  1\right)  & K_{2,1}\left(  1\right)  & K_{2,2}\left(  0\right)
\end{array}
\right\vert .
\end{align*}
But this is the same matrix as above.

\textbf{2.} Now we consider the case $\psi_{1}\left(  u\right)  =\left(
1-\alpha_{1}u\right)  ^{-1},$ $\psi_{2}\left(  u\right)  =\left(  1-\alpha
_{2}u\right)  ^{-1}.$ We have to prove that
\begin{equation}
\alpha_{1}\mathbf{P{}r}_{K}\left\{  \left(
\begin{array}
[c]{ccc}%
0 & \ast & \ast\\
\ast & 1 & 0
\end{array}
\right)  _{0,1,2}\right\}  =\alpha_{2}\mathbf{P{}r}_{K}\left\{  \left(
\begin{array}
[c]{ccc}%
0 & 1 & \ast\\
\ast & \ast & 0
\end{array}
\right)  _{0,1,2}\right\}  , \label{78}%
\end{equation}
or%
\begin{align*}
&  \alpha_{1}\det\left\vert
\begin{array}
[c]{ccc}%
K_{0,0}\left(  0\right)  -1 & K_{0,1}\left(  1\right)  & K_{0,2}\left(
1\right) \\
K_{1,0}\left(  -1\right)  & K_{1,1}\left(  0\right)  & K_{1,2}\left(  0\right)
\\
K_{2,0}\left(  -1\right)  & K_{2,1}\left(  0\right)  & K_{2,2}\left(
0\right)  -1
\end{array}
\right\vert \\
&  =\alpha_{2}\det\left\vert
\begin{array}
[c]{ccc}%
K_{0,0}\left(  0\right)  -1 & K_{0,1}\left(  0\right)  & K_{0,2}\left(
1\right) \\
K_{1,0}\left(  0\right)  & K_{1,1}\left(  0\right)  & K_{1,2}\left(  1\right)
\\
K_{2,0}\left(  -1\right)  & K_{2,1}\left(  -1\right)  & K_{2,2}\left(
0\right)  -1
\end{array}
\right\vert .
\end{align*}

Applying the relation $\left(  \ref{90}\right)  $ to the first two rows of the
first determinant, we see that
\begin{align*}
&  \det\left\vert
\begin{array}
[c]{ccc}%
K_{0,0}\left(  0\right)  -1 & K_{0,1}\left(  1\right)  & K_{0,2}\left(
1\right) \\
\alpha_{1}K_{1,0}\left(  -1\right)  & \alpha_{1}K_{1,1}\left(  0\right)  &
\alpha_{1}K_{1,2}\left(  0\right) \\
K_{2,0}\left(  -1\right)  & K_{2,1}\left(  0\right)  & K_{2,2}\left(
0\right)  -1
\end{array}
\right\vert \\
&  =\det\left\vert
\begin{array}
[c]{ccc}%
K_{0,0}\left(  0\right)  -1 & K_{0,1}\left(  1\right)  & K_{0,2}\left(
1\right) \\
K_{1,0}\left(  0\right)  & K_{1,1}\left(  1\right)  & K_{1,2}\left(  1\right)
\\
K_{2,0}\left(  -1\right)  & K_{2,1}\left(  0\right)  & K_{2,2}\left(
0\right)  -1
\end{array}
\right\vert .
\end{align*}
Applying now the relation $\left(  \ref{88}\right)  $ to the second and third
columns of the second determinant, we see the same result:%

\begin{align*}
&  \det\left\vert
\begin{array}
[c]{ccc}%
K_{0,0}\left(  0\right)  -1 & \alpha_{2}K_{0,1}\left(  0\right)  &
K_{0,2}\left(  1\right) \\
K_{1,0}\left(  0\right)  & \alpha_{2}K_{1,1}\left(  0\right)  & K_{1,2}\left(
1\right) \\
K_{2,0}\left(  -1\right)  & \alpha_{2}K_{2,1}\left(  -1\right)  &
K_{2,2}\left(  0\right)  -1
\end{array}
\right\vert \\
&  =\det\left\vert
\begin{array}
[c]{ccc}%
K_{0,0}\left(  0\right)  -1 & K_{0,1}\left(  1\right)  & K_{0,2}\left(
1\right) \\
K_{1,0}\left(  0\right)  & K_{1,1}\left(  1\right)  & K_{1,2}\left(  1\right)
\\
K_{2,0}\left(  -1\right)  & K_{2,1}\left(  0\right)  & K_{2,2}\left(
0\right)  -1
\end{array}
\right\vert .
\end{align*}

\textbf{3. }We go to the case $\psi_{1}\left(  u\right)  =\left(  1+\beta
_{1}u\right)  ,$ $\psi_{2}\left(  u\right)  =\left(  1-\alpha_{2}u\right)
^{-1}.$ Here we need to see that%
\begin{equation}
\beta_{1}\mathbf{P{}r}_{K}\left\{  \left(
\begin{array}
[c]{ccc}%
\ast & \ast & \ast\\
1 & 1 & 0
\end{array}
\right)  _{0,1,2}\right\}  =\alpha_{2}\mathbf{P{}r}_{K}\left\{  \left(
\begin{array}
[c]{ccc}%
\ast & 1 & \ast\\
1 & \ast & 0
\end{array}
\right)  _{0,1,2}\right\}  , \label{77}%
\end{equation}
which is the same as%
\[
\beta_{1}\mathbf{P{}r}_{K}\left\{  \left(
\begin{array}
[c]{ccc}%
\ast & 0 & \ast\\
1 & \ast & 0
\end{array}
\right)  _{0,1,2}\right\}  =\alpha_{2}\mathbf{P{}r}_{K}\left\{  \left(
\begin{array}
[c]{ccc}%
\ast & 1 & \ast\\
1 & \ast & 0
\end{array}
\right)  _{0,1,2}\right\}  .
\]
(The equivalence of the two identities follows from the simplest
case of the interlacing property proved in the previous section.)

Expressed via determinants, this is the relation%
\begin{align*}
&  \beta_{1}\det\left\vert
\begin{array}
[c]{ccc}%
K_{0,0}\left(  0\right)  & K_{0,1}\left(  -1\right)  & K_{0,2}\left(  0\right)
\\
K_{1,0}\left(  1\right)  & K_{1,1}\left(  0\right)  -1 & K_{1,2}\left(
1\right) \\
K_{2,0}\left(  0\right)  & K_{2,1}\left(  -1\right)  & K_{2,2}\left(
0\right)  -1
\end{array}
\right\vert \\
&  =-\alpha_{2}\det\left\vert
\begin{array}
[c]{ccc}%
K_{0,0}\left(  0\right)  & K_{0,1}\left(  -1\right)  & K_{0,2}\left(  0\right)
\\
K_{1,0}\left(  1\right)  & K_{1,1}\left(  0\right)  & K_{1,2}\left(  1\right)
\\
K_{2,0}\left(  0\right)  & K_{2,1}\left(  -1\right)  & K_{2,2}\left(
0\right)  -1
\end{array}
\right\vert .
\end{align*}
By $\left(  \ref{81}\right)  ,$ subtracting in the first determinant,%
\[
\det\left\vert
\begin{array}
[c]{ccc}%
K_{0,0}\left(  0\right)  & \beta_{1}K_{0,1}\left(  -1\right)  & K_{0,2}\left(
0\right) \\
K_{1,0}\left(  1\right)  & \beta_{1}\left[  K_{1,1}\left(  0\right)  -1\right]
& K_{1,2}\left(  1\right) \\
K_{2,0}\left(  0\right)  & \beta_{1}K_{2,1}\left(  -1\right)  & K_{2,2}\left(
0\right)  -1
\end{array}
\right\vert
\]
the first column from the second one, makes it into%
\[
\det\left\vert
\begin{array}
[c]{ccc}%
K_{0,0}\left(  0\right)  & -K_{0,1}\left(  0\right)  & K_{0,2}\left(  0\right)
\\
K_{1,0}\left(  1\right)  & -K_{1,1}\left(  1\right)  & K_{1,2}\left(  1\right)
\\
K_{2,0}\left(  0\right)  & -K_{2,1}\left(  0\right)  & K_{2,2}\left(
0\right)  -1
\end{array}
\right\vert .
\]

From $\left(  \ref{88}\right)  ,$ adding in the second determinant,%
\[
-\det\left\vert
\begin{array}
[c]{ccc}%
K_{0,0}\left(  0\right)  & \alpha_{2}K_{0,1}\left(  -1\right)  &
K_{0,2}\left(  0\right) \\
K_{1,0}\left(  1\right)  & \alpha_{2}K_{1,1}\left(  0\right)  & K_{1,2}\left(
1\right) \\
K_{2,0}\left(  0\right)  & \alpha_{2}K_{2,1}\left(  -1\right)  &
K_{2,2}\left(  0\right)  -1
\end{array}
\right\vert ,
\]
the third column to the second one results in%
\[
-\det\left\vert
\begin{array}
[c]{ccc}%
K_{0,0}\left(  0\right)  & K_{0,1}\left(  0\right)  & K_{0,2}\left(  0\right)
\\
K_{1,0}\left(  1\right)  & K_{1,1}\left(  1\right)  & K_{1,2}\left(  1\right)
\\
K_{2,0}\left(  0\right)  & K_{2,1}\left(  0\right)  & K_{2,2}\left(  0\right)
-1
\end{array}
\right\vert ,
\]
which is what we need.

\textbf{4. }The remaining case is $\psi_{1}\left(  u\right)  =\left(
1-\alpha_{1}u\right)  ^{-1},$ $\psi_{2}\left(  u\right)  =\left(  1+\beta
_{2}u\right)  .$ Here we need to see that%
\begin{equation}
\alpha_{1}\mathbf{P{}r}_{K}\left\{  \left(
\begin{array}
[c]{ccc}%
0 & \ast & 1\\
\ast & 1 & \ast
\end{array}
\right)  _{0,1,2}\right\}  =\beta_{2}\mathbf{P{}r}_{K}\left\{  \left(
\begin{array}
[c]{ccc}%
0 & 1 & 1\\
\ast & \ast & \ast
\end{array}
\right)  _{0,1,2}\right\}  , \label{72}%
\end{equation}
which is the same as%
\[
\alpha_{1}\mathbf{P{}r}_{K}\left\{  \left(
\begin{array}
[c]{ccc}%
0 & \ast & 1\\
\ast & 1 & \ast
\end{array}
\right)  _{0,1,2}\right\}  =\beta_{2}\mathbf{P{}r}_{K}\left\{  \left(
\begin{array}
[c]{ccc}%
0 & \ast & 1\\
\ast & 0 & \ast
\end{array}
\right)  _{0,1,2}\right\}  .
\]
The determinant relation to be checked is%
\begin{align*}
&  -\alpha_{1}\det\left\vert
\begin{array}
[c]{ccc}%
K_{0,0}\left(  0\right)  -1 & K_{0,1}\left(  1\right)  & K_{0,2}\left(
0\right) \\
K_{1,0}\left(  -1\right)  & K_{1,1}\left(  0\right)  & K_{1,2}\left(
-1\right) \\
K_{2,0}\left(  0\right)  & K_{2,1}\left(  1\right)  & K_{2,2}\left(  0\right)
\end{array}
\right\vert \\
&  =\beta_{2}\det\left\vert
\begin{array}
[c]{ccc}%
K_{0,0}\left(  0\right)  -1 & K_{0,1}\left(  1\right)  & K_{0,2}\left(
0\right) \\
K_{1,0}\left(  -1\right)  & K_{1,1}\left(  0\right)  -1 & K_{1,2}\left(
-1\right) \\
K_{2,0}\left(  0\right)  & K_{2,1}\left(  1\right)  & K_{2,2}\left(  0\right)
\end{array}
\right\vert .
\end{align*}

By $\left(  \ref{90}\right)  ,$ applied to the first determinant,%
\[
-\det\left\vert
\begin{array}
[c]{ccc}%
K_{0,0}\left(  0\right)  -1 & K_{0,1}\left(  1\right)  & K_{0,2}\left(
0\right) \\
\alpha_{1}K_{1,0}\left(  -1\right)  & \alpha_{1}K_{1,1}\left(  0\right)  &
\alpha_{1}K_{1,2}\left(  -1\right) \\
K_{2,0}\left(  0\right)  & K_{2,1}\left(  1\right)  & K_{2,2}\left(  0\right)
\end{array}
\right\vert ,
\]
the addition of the first row to the second one makes it into%
\[
-\det\left\vert
\begin{array}
[c]{ccc}%
K_{0,0}\left(  0\right)  -1 & K_{0,1}\left(  1\right)  & K_{0,2}\left(
0\right) \\
K_{1,0}\left(  0\right)  & K_{1,1}\left(  1\right)  & K_{1,2}\left(  0\right)
\\
K_{2,0}\left(  0\right)  & K_{2,1}\left(  1\right)  & K_{2,2}\left(  0\right)
\end{array}
\right\vert .
\]
Applying $\left(  \ref{82}\right)  $ to the determinant
\[
\det\left\vert
\begin{array}
[c]{ccc}%
K_{0,0}\left(  0\right)  -1 & K_{0,1}\left(  1\right)  & K_{0,2}\left(
0\right) \\
\beta_{2}K_{1,0}\left(  -1\right)  & \beta_{2}K_{1,1}\left(  0\right)  -1 &
\beta_{2}K_{1,2}\left(  -1\right) \\
K_{2,0}\left(  0\right)  & K_{2,1}\left(  1\right)  & K_{2,2}\left(  0\right)
\end{array}
\right\vert ,
\]
we turn it, after subtracting the third row from the second one, into%
\[
\det\left\vert
\begin{array}
[c]{ccc}%
K_{0,0}\left(  0\right)  -1 & K_{0,1}\left(  1\right)  & K_{0,2}\left(
0\right) \\
-K_{1,0}\left(  0\right)  & -K_{1,1}\left(  1\right)  & -K_{1,2}\left(
0\right) \\
K_{2,0}\left(  0\right)  & K_{2,1}\left(  1\right)  & K_{2,2}\left(  0\right)
\end{array}
\right\vert .
\]
That finishes our proof.

\subsection{\label{gmoves} Moves in general environment}

Now we have to check that the identities of the previous subsection holds in
more general situation. For the $\beta\beta$ case it means for example that%

\[
\beta_{1}\mathbf{P{}r}_{K}\left\{  \left(
\begin{array}
[c]{ccc}%
\ast & 0 & 1\\
1 & 1 & \ast
\end{array}
\right)  _{0,1,2}\cup n_{V}\right\}  =\beta_{2}\mathbf{P{}r}_{K}\left\{
\left(
\begin{array}
[c]{ccc}%
\ast & 1 & 1\\
1 & 0 & \ast
\end{array}
\right)  _{0,1,2}\cup n_{V}\right\}  ,
\]
where $V\subset\mathbb{Z}^{2}$ is an arbitrary finite set, disjoint
from the set $\left(
\begin{array}
[c]{ccc}
& \ast & \ast\\
\ast & \ast &
\end{array}
\right)  _{0,1,2},$ and the symbol $\left\{  \left(
\begin{array}
[c]{ccc}%
\ast & 0 & 1\\
1 & 1 & \ast
\end{array}
\right)  _{0,1,2}\cup n_{V}\right\}  $ denotes the event that we have the
configuration $\left(
\begin{array}
[c]{ccc}%
\ast & 0 & 1\\
1 & 1 & \ast
\end{array}
\right)  _{0,1,2}$ in our window, and all the sites in $V$ are occupied by the
particles. Consider the case when $V$ is just a single site $\left(
\zeta,z\right)  \in\mathbb{Z}^{2}.$ Let us see that the same relations which
were used in the subsection \ref{emoves}, work here as well.

We need to show that%
\begin{align}
&  \beta_{1}\det\left\vert
\begin{array}
[c]{ccccc}%
K_{0,0}\left(  0\right)   & K_{0,1}\left(  0\right)   & K_{0,1}\left(
-1\right)   & K_{0,2}\left(  -1\right)   & K_{0,\zeta}\left(  -z\right)  \\
K_{1,0}\left(  0\right)   & K_{1,1}\left(  0\right)   & K_{1,1}\left(
-1\right)   & K_{1,2}\left(  -1\right)   & K_{1,\zeta}\left(  -z\right)  \\
K_{1,0}\left(  1\right)   & K_{1,1}\left(  1\right)   & K_{1,1}\left(
0\right)  -1 & K_{1,2}\left(  0\right)   & K_{1,\zeta}\left(  1-z\right)  \\
K_{2,0}\left(  1\right)   & K_{2,1}\left(  1\right)   & K_{2,1}\left(
0\right)   & K_{2,2}\left(  0\right)   & K_{2,\zeta}\left(  1-z\right)  \\
K_{\zeta,0}\left(  z\right)   & K_{\zeta,1}\left(  z\right)   & K_{\zeta
,1}\left(  z-1\right)   & K_{\zeta,2}\left(  z-1\right)   & K_{\zeta,\zeta
}\left(  0\right)
\end{array}
\right\vert \label{62}\\
&  =\beta_{2}\det\left\vert
\begin{array}
[c]{ccccc}%
K_{0,0}\left(  0\right)   & K_{0,1}\left(  0\right)   & K_{0,1}\left(
-1\right)   & K_{0,2}\left(  -1\right)   & K_{0,\zeta}\left(  -z\right)  \\
K_{1,0}\left(  0\right)   & K_{1,1}\left(  0\right)  -1 & K_{1,1}\left(
-1\right)   & K_{1,2}\left(  -1\right)   & K_{1,\zeta}\left(  -z\right)  \\
K_{1,0}\left(  1\right)   & K_{1,1}\left(  1\right)   & K_{1,1}\left(
0\right)   & K_{1,2}\left(  0\right)   & K_{1,\zeta}\left(  1-z\right)  \\
K_{2,0}\left(  1\right)   & K_{2,1}\left(  1\right)   & K_{2,1}\left(
0\right)   & K_{2,2}\left(  0\right)   & K_{2,\zeta}\left(  1-z\right)  \\
K_{\zeta,0}\left(  z\right)   & K_{\zeta,1}\left(  z\right)   & K_{\zeta
,1}\left(  z-1\right)   & K_{\zeta,2}\left(  z-1\right)   & K_{\zeta,\zeta
}\left(  0\right)
\end{array}
\right\vert .\nonumber
\end{align}
It is immediate to see that the same strategy which was used in the simplest
case $4\times4$ works: application of $\left(  \ref{81}\right)  $ turns the
determinant%
\[
\det\left\vert
\begin{array}
[c]{ccccc}%
K_{0,0}\left(  0\right)   & K_{0,1}\left(  0\right)   & \beta_{1}%
K_{0,1}\left(  -1\right)   & K_{0,2}\left(  -1\right)   & K_{0,\zeta}\left(
-z\right)  \\
K_{1,0}\left(  0\right)   & K_{1,1}\left(  0\right)   & \beta_{1}%
K_{1,1}\left(  -1\right)   & K_{1,2}\left(  -1\right)   & K_{1,\zeta}\left(
-z\right)  \\
K_{1,0}\left(  1\right)   & K_{1,1}\left(  1\right)   & \beta_{1}\left[
K_{1,1}\left(  0\right)  -1\right]   & K_{1,2}\left(  0\right)   & K_{1,\zeta
}\left(  1-z\right)  \\
K_{2,0}\left(  1\right)   & K_{2,1}\left(  1\right)   & \beta_{1}%
K_{2,1}\left(  0\right)   & K_{2,2}\left(  0\right)   & K_{2,\zeta}\left(
1-z\right)  \\
K_{\zeta,0}\left(  z\right)   & K_{\zeta,1}\left(  z\right)   & \beta
_{1}K_{\zeta,1}\left(  z-1\right)   & K_{\zeta,2}\left(  z-1\right)   &
K_{\zeta,\zeta}\left(  0\right)
\end{array}
\right\vert
\]
into%
\[
\det\left\vert
\begin{array}
[c]{ccccc}%
K_{0,0}\left(  0\right)   & K_{0,1}\left(  0\right)   & 0 & K_{0,2}\left(
-1\right)   & K_{0,\zeta}\left(  -z\right)  \\
K_{1,0}\left(  0\right)   & K_{1,1}\left(  0\right)   & 1 & K_{1,2}\left(
-1\right)   & K_{1,\zeta}\left(  -z\right)  \\
K_{1,0}\left(  1\right)   & K_{1,1}\left(  1\right)   & 0 & K_{1,2}\left(
0\right)   & K_{1,\zeta}\left(  1-z\right)  \\
K_{2,0}\left(  1\right)   & K_{2,1}\left(  1\right)   & 0 & K_{2,2}\left(
0\right)   & K_{2,\zeta}\left(  1-z\right)  \\
K_{\zeta,0}\left(  z\right)   & K_{\zeta,1}\left(  z\right)   & 0 &
K_{\zeta,2}\left(  z-1\right)   & K_{\zeta,\zeta}\left(  0\right)
\end{array}
\right\vert ,
\]
while the rhs of $\left(  \ref{62}\right)  $ is treated as the rhs of $\left(
\ref{61}\right)  .$ So one sees in this way that the identities of the
subsection \ref{emoves} work for all sets $V$ in all the cases.

\subsection{Interlacing property. General case.}

\textbf{1. }The case $\psi_{1}\left(  u\right)  =\left(  1-\alpha_{1}u\right)
^{-1}.$ We will show now that
\begin{align}
&  \mathbf{P{}r}_{K}\left\{  \left(
\begin{array}
[c]{cc}%
1 & \ast\\
0 & \ast\\
... & \ast\\
0 & \ast\\
1 & \ast
\end{array}
\right)  _{0,1}\right\}  =\mathbf{P{}r}_{K}\left\{  \left(
\begin{array}
[c]{cc}%
1 & \ast\\
0 & 0\\
... & ...\\
0 & 0\\
1 & 1
\end{array}
\right)  _{0,1}\right\}  +\label{80}\\
&  +\mathbf{P{}r}_{K}\left\{  \left(
\begin{array}
[c]{cc}%
1 & \ast\\
0 & 0\\
... & ...\\
0 & 1\\
1 & 0
\end{array}
\right)  _{0,1}\right\}  +...+\mathbf{P{}r}_{K}\left\{  \left(
\begin{array}
[c]{cc}%
1 & \ast\\
0 & 1\\
... & ...\\
0 & 0\\
1 & 0
\end{array}
\right)  _{0,1}\right\}  ,\nonumber
\end{align}
that is, if we have a configuration with two particles, separated by
a vertical string of $n-2$ holes, then with probability one the next
column to the right has in the lower $n-1$ cells exactly one
particle and $n-2$ holes. (So our sum above has $n-1$ terms.) This
is the particle interlacing property. We will prove it
simultaneously with the complementary (reflected) statement:%
\begin{align}
&  \mathbf{P{}r}_{K}\left\{  \left(
\begin{array}
[c]{cc}%
\ast & 1\\
\ast & 0\\
... & ...\\
\ast & 0\\
\ast & 1
\end{array}
\right)  _{0,1}\right\}  =\mathbf{P{}r}_{K}\left\{  \left(
\begin{array}
[c]{cc}%
0 & 1\\
0 & 0\\
... & ...\\
1 & 0\\
\ast & 1
\end{array}
\right)  _{0,1}\right\}  +\label{79}\\
&  +...+\mathbf{P{}r}_{K}\left\{  \left(
\begin{array}
[c]{cc}%
0 & 1\\
1 & 0\\
... & ...\\
0 & 0\\
\ast & 1
\end{array}
\right)  _{0,1}\right\}  +\mathbf{P{}r}_{K}\left\{  \left(
\begin{array}
[c]{cc}%
1 & 1\\
0 & 0\\
... & ...\\
0 & 0\\
\ast & 1
\end{array}
\right)  _{0,1}\right\}  .\nonumber
\end{align}
The proof goes by induction on the length of the aforementioned
string of the holes. The case of the empty string -- i.e. $n=2$ --
was dealt with in Section \ref{support}. So suppose that we know
already the relations $\left( \ref{80}\right)  $ and $\left(
\ref{79}\right)  $ for all $n<k.$ Let us prove them for $n=k.$

First we can exclude the case in $\left(  \ref{80}\right)  $ when we have at
least two particles in the second column. Indeed, that means that we have
there a pattern $\left(
\begin{array}
[c]{cc}%
0 & 1\\
0 & 0\\
... & ...\\
0 & 0\\
\ast & 1
\end{array}
\right)  _{0,1}$ with the string of holes in the second column of length less
than $k-2,$ which is ruled out by induction hypothesis for $\left(
\ref{79}\right)  .$ The same argument applies to $\left(  \ref{79}\right)  .$

It remains to show that $\mathbf{P{}r}_{K}\left\{  \left(
\begin{array}
[c]{cc}%
1 & \ast\\
0 & 0\\
... & ...\\
0 & 0\\
1 & 0
\end{array}
\right)  _{0,1}\right\}  =0,$ where we have $k-1$ holes in the right column.
Here we note that the above probability depends on $K$ only through one
parameter, $\alpha_{1}.$ So without loss of generality we can assume that we
are in the \textquotedblleft$\alpha\alpha$\textquotedblright\ situation, i.e.
that $\psi_{0}\left(  u\right)  =\left(  1-\alpha_{0}u\right)  ^{-1}.$Let us
write our event as a sum of four events:%
\begin{align}
\left(
\begin{array}
[c]{cc}%
1 & \ast\\
0 & 0\\
... & ...\\
0 & 0\\
1 & 0
\end{array}
\right)  _{0,1}  &  =\left(
\begin{array}
[c]{ccc}%
1 & 1 & \ast\\
\ast & 0 & 0\\
\ast & ... & ...\\
1 & 0 & 0\\
\ast & 1 & 0
\end{array}
\right)  _{-1,0,1}+\left(
\begin{array}
[c]{ccc}%
1 & 1 & \ast\\
\ast & 0 & 0\\
\ast & ... & ...\\
0 & 0 & 0\\
\ast & 1 & 0
\end{array}
\right)  _{-1,0,1}\label{76}\\
&  +\left(
\begin{array}
[c]{ccc}%
0 & 1 & \ast\\
\ast & 0 & 0\\
\ast & ... & ...\\
1 & 0 & 0\\
\ast & 1 & 0
\end{array}
\right)  _{-1,0,1}+\left(
\begin{array}
[c]{ccc}%
0 & 1 & \ast\\
\ast & 0 & 0\\
\ast & ... & ...\\
0 & 0 & 0\\
\ast & 1 & 0
\end{array}
\right)  _{-1,0,1}.\nonumber
\end{align}
The first one has zero probability; this is our induction assumption. And to
every one of the remaining events we can apply the move transformation
$\left(  \ref{78}\right)  ,$ which makes the two particles in the middle
column to become one unit closer. After that we get a configuration, which has
zero probability by induction hypothesis. This ends the proof of our statement.

\textbf{2. }The case $\psi_{1}\left(  u\right)  =\left(  1+\beta_{1}u\right)
^{-1}.$ We will show that%
\begin{align}
&  \mathbf{P{}r}_{K}\left\{  \left(
\begin{array}
[c]{cc}%
0 & \ast\\
1 & \ast\\
... & \ast\\
1 & \ast\\
0 & \ast
\end{array}
\right)  _{0,1}\right\}  =\mathbf{P{}r}_{K}\left\{  \left(
\begin{array}
[c]{cc}%
0 & 0\\
1 & 1\\
... & ...\\
1 & 1\\
0 & \ast
\end{array}
\right)  _{0,1}\right\}  +\label{73}\\
&  +\mathbf{P{}r}_{K}\left\{  \left(
\begin{array}
[c]{cc}%
0 & 1\\
1 & 0\\
... & ...\\
1 & 1\\
0 & \ast
\end{array}
\right)  _{0,1}\right\}  +...+\mathbf{P{}r}_{K}\left\{  \left(
\begin{array}
[c]{cc}%
0 & 1\\
1 & 1\\
... & ...\\
1 & 0\\
0 & \ast
\end{array}
\right)  _{0,1}\right\}  .\nonumber
\end{align}
Here in the lhs we have the probability of the event that two holes
are separated by the string of $n-2$ particles, while in the rhs we
have a sum of probabilities of the $n-1$ events that the right
column has exactly one hole in the upper $n-1$ positions. This is
the hole interlacing. Again, we will prove it by induction on $n,$
the case $n=2$ was established above, see $\left(  \ref{75}\right)
.$ We will treat simultaneously the reflected event
as well (compare with $\left(  \ref{80}\right)  $ and $\left(  \ref{79}%
\right)  $.)

The presence of more than one hole in the right column in $\left(
\ref{73}\right)  $ is again ruled out by induction. To study the probability
of the event $\left(
\begin{array}
[c]{cc}%
0 & 1\\
1 & 1\\
... & ...\\
1 & 1\\
0 & \ast
\end{array}
\right)  _{0,1}$ we can without loss of generality consider the case $\psi
_{0}\left(  u\right)  =\left(  1+\beta_{0}u\right)  ^{-1}$, and we write%
\begin{align*}
\left(
\begin{array}
[c]{cc}%
0 & 1\\
1 & 1\\
... & ...\\
1 & 1\\
0 & \ast
\end{array}
\right)  _{0,1}  &  =\left(
\begin{array}
[c]{ccc}%
\ast & 0 & 1\\
0 & 1 & 1\\
\ast & ... & ...\\
\ast & 1 & 1\\
0 & 0 & \ast
\end{array}
\right)  _{-1,0,1}+\left(
\begin{array}
[c]{ccc}%
\ast & 0 & 1\\
1 & 1 & 1\\
\ast & ... & ...\\
\ast & 1 & 1\\
0 & 0 & \ast
\end{array}
\right)  _{-1,0,1}\\
&  +\left(
\begin{array}
[c]{ccc}%
\ast & 0 & 1\\
0 & 1 & 1\\
\ast & ... & ...\\
\ast & 1 & 1\\
1 & 0 & \ast
\end{array}
\right)  _{-1,0,1}+\left(
\begin{array}
[c]{ccc}%
\ast & 0 & 1\\
1 & 1 & 1\\
\ast & ... & ...\\
\ast & 1 & 1\\
1 & 0 & \ast
\end{array}
\right)  _{-1,0,1}.
\end{align*}
The first event is ruled out by induction, while the remaining three
are movable, and the application of the corresponding move (the
first one described in Section \ref{emoves}) reduces the length of
the string of particles in the middle column ($0$-th one) by one, so
the remaining three events also have vanishing probability.

\subsection{The downward paths.}

We will show now that the case of the functions $\psi$-s of the types $\left(
1-\alpha_{k}^{-}u^{-1}\right)  ^{-1}$ and $\left(  1+\beta_{k}^{-}%
u^{-1}\right)  $ can be reduced to the one when all $\psi_{k}\left(  u\right)
$ are of the form $\left(  1-\alpha_{k}^{+}u\right)  ^{-1}$ or $\left(
1+\beta_{k}^{+}u\right)  .$ Indeed, we have the identities
\[
\left(  1-\alpha_{k}^{-}u^{-1}\right)  ^{-1}=-\alpha_{k}^{-}u\left(
1-(\alpha_{k}^{-})^{-1}u\right)  ^{-1},
\]%
\[
\left(  1+\beta_{k}^{-}u^{-1}\right)  =\beta_{k}^{-}u^{-1}\left(
1+(\beta _{k}^{-})^{-1}u\right)  .
\]

Observe that multiplication of $\psi_{k}(u)$ by a constant $c$ leads to the
conjugation of the kernel:
\[
K(\sigma,x;\tau,y)\mapsto\left\{
\begin{array}
[c]{cc}%
cK(\sigma,x;\tau,y) & \text{if }\sigma\geq k>\tau,\\
c^{-1}K(\sigma,x;\tau,y) & \text{if }\tau\geq k>\sigma,\\
K(\sigma,x;\tau,y) & \text{ otherwise},
\end{array}
\right.  \newline\newline%
\]
which does not affect the determinants for the correlation functions. On the
other hand, the multiplication of $\psi_{k}(u)$ by $u$ in the formula for the
kernel is equivalent to the following transformation of the state space
$\mathbb{Z}^{2}$:
\[
(\sigma,x)\mapsto\left\{
\begin{array}
[c]{cc}%
(\sigma,x) & \text{ if }\sigma<k,\\
(\sigma,x+1) & \text{ if }\sigma\geq k.
\end{array}
\right.
\]
Under this transformation every configuration which was satisfying
the downward interlacing (for particles or for holes) in the column
$\left\{ (\sigma,x):\sigma=k,k+1\right\}  $ will satisfy the upward
interlacing, so the new configuration can still be associated with a
collection of paths. It is straightforward to see that the change of
the weight of a path affected by elementary move is the same in both
path configurations, compare the definitions $\left( \ref{71}\right)
,$ $\left(  \ref{70}\right)  ,$ $\left( \ref{69}\right)  ,$ $\left(
\ref{68}\right)  .$ That proves our statement.

\subsection{The Gibbs property}

Now we are in the position to check the Gibbs property of the field $n$ viewed
as the probability distribution over the lattice paths built from the patterns
$\left(  \ref{66}\right)  ,$ $\left(  \ref{65}\right)  ,$ $\left(
\ref{64}\right)  $ and $\left(  \ref{63}\right)  .$ After the preliminary work
we did it is almost immediate.

Indeed, we have already checked in the subsections \ref{emoves},
\ref{gmoves} that the ratio of the probabilities of two
configurations $n_{V}^{\prime}$ and $n_{V}^{\prime\prime}$ which
differ by allowed move of one particle depends only on the type of
the move and equals to the exponent of the action difference for the
corresponding paths. Let us take, in particular, any (finite
simply-connected) box $\Lambda\subset \mathbb{Z}^{2},$ and fix the
sets $P$ and $Q$ of the entrance and exit points of the paths on the
boundary $\partial\Lambda$. Note that in that case any allowed
configuration of paths in $\Lambda$ can be obtained from any other
by a sequence of elementary moves. This claim is essentially
obvious; if follows from the fact that there is a minimal path
joining any two points (if the set of paths joining the two points
is nonempty), and induction on the number of paths. That finishes
our proof.

\section{Positivity}\label{positivity}

Denote by $\Om$ the set of elements $\varpi=(\al,\be,\ga)\in
\R_+^\infty\times\R_+^\infty\times \R_+$ such that
$$
\sum_{i=1}^\infty \al_i<\infty, \qquad \sum_{i=1}^\infty
\beta_i<\infty.
$$

For $\varpi\in \Om$, we denote by $\psi_{\varpi}$ the following
meromorphic functions on $\C$:
\begin{equation}\label{1}
\psi_{\varpi}(u)=e^{\gamma u}\prod_{j=1}^\infty \frac{1+\beta_j
u}{1-\alpha_j u}\,.
\end{equation}

For $\varpi^+,\varpi^-\in\Om$, we also set
$$
\psi_{\varpi^+,\varpi^-}(u)=\psi_{\varpi^+}(u)\psi_{\varpi^-}(u^{-1}).
$$
Coordinates of $\varpi^\pm$ will be denotes as
$\al^\pm_i,\beta^\pm_i,\ga^\pm$.

Our goal is to prove the following statement, which is a slight
generalization of Theorem 4.4 in \cite{B-Per}.

\begin{theorem} Fix a complex number $z$ with $\Im z>0$ and denote
$C_\pm$ any contour that joins $\bar{z}$ and $z$ and crosses the
real axis at a point of $\R_{\pm}$. Then for any doubly infinite
sequences ${\{\varpi^+[k],\varpi^-[k]\}}_{k\in\Z}$ of elements in
$\Om$, there exists a (unique) determinantal point process on
$\Z\times\Z$ with the correlation kernel
\begin{equation}
K(\sigma,x; \tau,y)=\begin{cases}\dfrac 1{2\pi
i}\displaystyle\int_{C_+} {\left(\prod_{k=\sigma+1}^\tau
\psi_{\varpi^+[k],\varpi^-[k]}(u) \right)^{-1}}\,
\dfrac{du}{u^{x-y+1}},&\sigma\le
\tau,\\
\dfrac 1{2\pi i}\displaystyle\int_{C_-} {\prod_{k=\tau+1}^\sigma
\psi_{\varpi^+[k],\varpi^-[k]}(u)}\,
\dfrac{du}{u^{x-y+1}},&\sigma>\tau.
\end{cases}
\label{2}
\end{equation}
\end{theorem}

\noindent {\bf Comments.} 1. The kernels considered in the previous
sections are the ones with each of
$\psi_{\varpi^+[k],\varpi^-[k]}(u)$ having the form either
$(1-\alpha_k u)^{-1}$ or $(1+\beta_k u)$.
\medskip

\noindent 2. The classical fact that lies at the foundation of this
theorem is that functions $\psi_\varpi(u)$ are generating functions
of the totally positive sequences. This statement was independently
proved by Aissen-Edrei-Schoenberg-Whitney in 1951 \cite{AESW},
\cite{Ed}, and by Thoma in 1964 \cite{Th}. An excellent exposition
of deep relations of this result to representation theory of the
infinite symmetric group can be found in Kerov's book \cite{Ke}.

\medskip

\noindent 3. The equal time restriction of the kernel above is
equivalent to the discrete sine kernel on $\Z$; for any $\tau\in\Z$
$$
K(\tau,x;\tau,y)=\dfrac 1{2\pi i}\int_{C_+}
\dfrac{d\zeta}{\zeta^{x-y+1}}=\frac{e^{|z|
y}}{e^{|z|x}}\,\frac{\sin((\arg z)(x-y))}{\pi (x-y)}\,.
$$
In particular, the density of particles is equal to $\arg z/\pi$
everywhere. The kernels $K(\sigma,x;\tau,y)$ may be viewed as
extensions of the discrete sine kernel.

\medskip

\noindent 4. The class of the random point processes afforded by
this theorem is closed under

$\bullet$ projections of $\Z\times\Z$ to $A\times \Z$, where
$A=\{a_n\}_{n=-\infty}^{+\infty}$ is any doubly infinite sequence of
integers;

$\bullet$ shifts and reflection of either of the two coordinates of
$\Z\times\Z$;

$\bullet$ particle-hole inversion on any subset of the form $B\times
\Z$, where $B\subset\Z$.

\medskip

\noindent 5. The projection of the process to the set
$\{1,\dots,T\}\times \Z$ depend only on $\varpi^\pm[k]$ with
$k=1,\dots,T$.

\medskip

We will give two proofs of the theorem; one is essentially a
reduction to Theorem 4.4 from \cite{B-Per}, while the second one is
``more constructive'' --- it explains how to build our process from
a deformation of the uniform measure on large plane partitions.

\medskip

\noindent \emph{Proof 1.} Observe that the change of the integration
variable $u = rv$, $r>0$, replaces the formula for the kernel by a
similar one with $z\mapsto z/r$, all coordinates of $\varpi^+_k$'s
multiplied by $r$, all coordinates of $\varpi^-_k$'s divided by $r$,
and the integral itself multiplied by $r^{y-x}$. The prefactor
$r^{y-x}$ cancels out in the determinants of the form $\det
[K(\sigma_i,x_i; \sigma_j,x_j)]$, thus it can be removed. Hence, it
suffices to prove the claim for $z$ with $|z|=1$.

Observe further, that multiplication of
$\psi_{\varpi^+[m],\varpi^-[m]}(u)$ by $u^n$ in the formula for the
kernel above is equivalent to the following transformation of the
state space $\Z\times\Z$:
$$
(\sigma,x)\mapsto\begin{cases} (\sigma,x),&\sigma<
m,\\(\sigma,x+n),&\sigma\ge m. \end{cases}
$$
On the other hand, multiplication of
$\psi_{\varpi^+[m],\varpi^-[m]}(u)$ by a constant $c$ leads to the
conjugation of the kernel
$$
K(\sigma,x; \tau,y)\mapsto \begin{cases} c K(\sigma,x;
\tau,y),&\sigma+1\le m\le \tau,\\
c^{-1} K(\sigma,x; \tau,y) ,&\tau+1\le m\le \sigma, \\
K(\sigma,x; \tau,y),&\text{otherwise}, \end{cases}
$$
which does not affect the determinants for the correlation
functions.

The identities
$$
1-\alpha u=-\alpha u\cdot(1-\alpha^{-1}u^{-1}),\qquad 1+\beta
u=\beta u\cdot(1+\beta^{-1}u^{-1})
$$
then show that we can freely replace parameters
$\alpha_i^+[k]=\alpha$ and $\beta_i^+[k]=\beta$ by
$\alpha_i^-[k]=\alpha^{-1}$ and $\beta_i^-[k]=\beta^{-1}$ and the
other way around, and such changes do not affect the statement that
the kernel defines a random point process.

Using such replacements we can then choose the parameters in such a
way that all $\alpha^\pm_i[k],\beta^\pm_i[k]$ are in the segment
$[0,1]$. Since the statement of the theorem is stable under limit
transitions, we can assume that all the parameters are strictly
smaller than 1 without loss of generality.

But if the parameters satisfy the conditions
$$
|z|=1,\qquad \alpha^\pm_i[k],\,\beta^\pm_i[k]<1, \quad \text{for all
$i$, $k$}
$$
then our claim is exactly Theorem 4.4 in \cite{B-Per}. $\square$

\medskip

\noindent \emph{Proof 2.} The argument is based on the Schur process
of \cite{OR} and can be constructed as follows. We use the
definitions and notation of \cite{OR}.

Let us construct a deformation of the Schur process. More precisely,
the Schur process is parameterized by two sequences
$\{\phi^+[m],\phi^-[m]\}_{m\in \Z+\frac 12}$ of functions
holomorphic and nonvanishing in some neighborhood of the interior
(resp., exterior) of the unit disc. In order for the process to
assign positive weights, the functions $\phi^\pm$ have to be such
that all minors of the triangular Toeplitz matrices with symbols
$\phi^+(u)$ and $\phi^-(u^{-1})$ are nonnegative; this is exactly
the content of Comment 2 above.

The concrete example of the Schur process studied asymptotically in
\cite{OR} corresponds to the choice
$$
\begin{gathered}
\phi^+[m](u)=\begin{cases} (1-q^{-m}u)^{-1},&m<0,\\
1,&m>0,\end{cases}\\ \phi^-[m](u)=\begin{cases} 1,&m<0,\\
(1-q^{m}u^{-1})^{-1},&m>0.\end{cases}
\end{gathered}
$$

Let us choose $N$ consecutive values of $m$, say
$M,M+1,\dots,M+N-1$, and replace the corresponding functions
$\phi^\pm$ as follows:
$$
\begin{gathered} \widetilde\phi^+[M+k](u)=\psi_{\varpi^+[k]}(u),\qquad
\widetilde\phi^-[M+k](u)=\psi_{\varpi^-[k]}(u^{-1}),
\end{gathered}
$$
for $k=0,\dots,N-1$.

Taking the point of the limit shape with $\tau=0$ (near the corner),
one readily sees that such a modification produces no impact on the
asymptotic analysis of \cite{OR} until the very last stage
--- the computation of the residue denoted as $\int^{(2)}$ in
Section 3.1.6.

The residue is an integral of the expression in formula (26) without
the factor $(z-w)$ in the denominator and with $z=w$, where
$\Phi(t,z)$ is defined by the formula (20). The computation gives
$$
\frac 1{2\pi i}\int_{\bar z_c}^{z_c}\prod_{m=t_j+1}^{t_i} \frac
1{\phi^+[m](w^{-1})\phi^-[m](w^{-1})}\,
\frac{dw}{w^{h_i-h_j+(t_i-t_j)/2+1}}
$$
for $t_i\ge t_j$ and
$$
-\frac 1{2\pi i}\int_{z_c}^{\bar{z}_c}\prod_{m=t_i+1}^{t_j}
{\phi^+[m](w^{-1})\phi^-[m](w^{-1})}\,\frac{dw}{w^{h_i-h_j+(t_i-t_j)/2+1}}
$$
for $t_i<t_j$. If we now choose $M$ in such a way that $t_i$ and
$t_j$ lie in the set $M,M+1,\dots,M+N-1$ then substituting the
deformed functions $\widetilde\phi^\pm[M+k](u)$ we arrive at the
kernel (\ref{2}) with the change of variables
$$
w\mapsto u^{-1},\quad z_c\mapsto z^{-1},\quad (t_i,t_j)\mapsto
(\tau,\sigma),\quad \left(h_i+\frac
{t_i}2,h_j+\frac{t_j}2\right)\mapsto (y,x).
$$

Thus, we showed that determinants made from the kernel (\ref{2}) are
limits of the correlation functions of certain point processes.
$\square$

\end{document}